\documentclass[12pt]{article}
\pdfoutput=1
\usepackage{amsthm}
\usepackage[margin=25truemm]{geometry}
\usepackage{multirow}
\usepackage[dvipdfmx]{graphicx}
\usepackage{tikz-feynhand}
\usepackage[utf8]{inputenc}
\usepackage{authblk}
\usepackage{MnSymbol}
\usepackage[colorlinks=true,linkcolor=magenta]{hyperref}
\usepackage{physics}
\usepackage{here}
\usepackage{mathtools}
\usepackage{bm}
\usepackage[version=3]{mhchem}
\usepackage[english]{babel}
\usepackage{color}
\usepackage{comment}
\usepackage[export]{adjustbox}
\usepackage[utf8]{inputenc}
\usepackage{color}

\usepackage{fancybox}
\usepackage{amsmath, amsfonts}

\newcommand{\mc}[1]{\mathcal{#1}}
\newcommand{\mr}[1]{\mathrm{#1}}

\newcommand{\mb}[1]{\mathbb{#1}}
\newcommand{\msf}[1]{\mathsf{#1}}

\newcommand{\normord}[1]{:\mathrel{#1}:}

\theoremstyle{remark}

\begin{document}
\title{Quantum many-body scars as remnants of stable many-body periodic orbits}
\author{Keita Omiya}
	\affil{PSI Center for Scientific Computing, Theory and Data, 5232 Villigen PSI, Switzerland}
	\affil{Institute of Physics, Ecole Polytechnique F\'ed\'erale de Lausanne (EPFL), CH-1015 Lausanne, Switzerland}
 \maketitle
		\begin{abstract}
			Quantum many-body scars (QMBS) represent a weak ergodicity-breaking phenomenon that defies the common scenario of thermalization in closed quantum systems. 
			They are often regarded as a many-body analog of quantum scars (QS)---a single-particle phenomenon in quantum chaos---due to their superficial similarities. 
			However, unlike QS, a clear connection between QMBS and classical chaos has remained elusive.
			It has nevertheless been speculated that in an appropriate semiclassical limit, QMBS should have a correspondance to weakly unstable periodic orbits. In this paper, 
			I present a counterexample to this conjecture by studying a bosonic model with a large number of flavors. The dynamics of out-of-time-ordered correlators (OTOCs) 
			suggest that QMBS do not display chaotic behavior in the semiclassical limit. In contrast, chaotic dynamics are expected for initial states not associated with QMBS. 
			Interestingly, the anomalous OTOC dynamics persist even under weak perturbations that eliminate the scarred eigenstates, suggesting a certain robustness in the phenomenon.
			   
	\end{abstract}

\section{Introduction and summary}
\renewcommand{\theequation}{1.\arabic{equation}}
\setcounter{equation}{0}

The question of thermalization in closed quantum many-body systems is one of the central problems in statistical physics. 
Although it remains an open problem requiring extensive study, the Eigenstate Thermalization 
Hypothesis (ETH)~\cite{eth_deutsch,eth_srednicki} is a widely accepted concept. The ETH posits that the expectation values evaluated 
in each eigenstate of a generic many-body system should coincide with that evaluated in the 
corresponding Gibbs ensemble in the thermodynamic limit. This hypothesis has been tested and confirmed primarily through numerical simulations across various systems, e.g., in Ref\,\cite{Rigol2008}.

However, Ref.\,\cite{mori_shiraishi} points out that thermalization does not necessarily imply the ETH. In fact, Ref.\,\cite{mori_shiraishi} outlines a systematic 
method of constructing non-integrable models which violate the ETH while still allowing for thermalization for almost all initial states, with the exception of 
a vanishing fraction of initial states that do not thermalize. 
This construction is known as the Shiraishi-Mori (SM) construction.
The core idea is that an ETH-violating subspace, characterized as the common kernel of local projectors, is embedded within the many-body Hilbert space, and the interaction is fine-tuned such that this subspace 
remains invariant under the Hamiltonian. The Hamiltonian from this construction typically takes the form $H=\sum_\alpha P_\alpha h_\alpha P_\alpha+H_Z$ with $P_\alpha$ a local projector characterizing the subspace 
and $H_Z$ a trivially integrable term (such as a Zeeman term) that preserves the special subspace. 
A straightforward consequence of the SM construction is that if the initial state is a superposition of eigenstates within the ETH-violating 
subspace, it will either (i) avoid thermalization or (ii) reach a stationary state that is not described by the standard Gibbs ensemble\,\cite{PhysRevE.96.022153}. 

A remarkable experiment involving an array of Rydberg atoms~\cite{qmbs_experiment} is believed to approximately realize the scenario (i). 
In this experiment, anomalous oscillations persisted much longer 
than the typical timescale of the system, but only when the initial state was the $\mb{Z}_2$-ordered state. A common interpretation of these oscillations is that the initial $\mb{Z}_2$ state is approximately 
a superposition of ETH-violating eigenstates, also known as scar states~\cite{Turner2018}, which possess an almost equidistant energy 
spectrum, yielding quasi-periodic motions that assures slow thermalization.

This phenomenon was termed quantum many-body scars (QMBS) in Ref.\,\cite{Turner2018} 
due to its resemblance to the single-particle phenomenon known as quantum scars (QS) in 
quantum billiards~\cite{quantum_scar_original}. A QS is a significant quantum correction to the 
Wigner function, leading to the concentration of certain atypical ``scarred'' eigenfunctions on classical unstable periodic orbits. 
This phenomenon provides a counterexample to Berry's conjecture asserting that the Wigner function of eigenstates in classically chaotic 
systems should be delocalized across the whole phase space allowed by symmetries. 
As a result, an initial Gaussian wavepacket also exhibits 
long-lasting oscillations, even in a chaotic billiard when placed on weakly unstable periodic orbits.  

Despite this loose analogy, the theory of QMBS has largely developed independently of the background known from QS. Numerous toy models, including the spin-1 XY model, have been proposed and analyzed, exhibiting 
similar phenomenology to the Rydberg experiment~\cite{tos_aklt,tos_xy,tos_exact_2}. Many of these toy models share common features, which led to several studies attempting a unified construction 
of those Hamiltonians~\cite{tos_aklt_unified,moudgalya2023}. One such feature is that the toy models can typically be decomposed into a Zeeman term and a sum of local terms, each of which annihilates the scar states. This 
structure was found to be a special case of the SM construction. Furthermore, even for the models previously thought to evade such an algebraic characterization, it has been shown that they can be cast into the form of the SM construction 
once the Hilbert space is properly chosen~\cite{PhysRevA.107.023318,PhysRevB.108.054412}.

While the algebraic aspects of QMBS have been intensively studied, their connection to QS and classical chaos is little understood, partly due to the lack of a parameter interpolating 
between the available quantum toy models and an appropriate classical limit. Such quantum models consist of quantum spins or fermions, which do not have well-defined classical limits. One possible approach to studying chaos in such systems 
is to use the time-dependent variational principle (TDVP), where the stability of a specific variational solution can be examined. Another approach is to move beyond such models and consider semiclassical systems, 
where chaos can be directly studied by solving the classical equation of motion. The former approach has been employed in the analysis of the PXP model~\cite{qmbs_tvdp_prl}, the effective model describing the Rydberg experiment, where the scar 
states are relatively well understood. However, this approach can only access the stability of the variational manifold, which may differ from that of the quantum state itself. The latter 
approach is more suited to studying chaos of the system~\cite{PhysRevLett.130.250402,PhysRevLett.132.020401,MM_2024}, but underlying scar states, if any, 
are usually difficult to obtain, and one typically relies on numerics that might suffer from finite-size effects. Nevertheless, a prevailing conjecture posits
that the proper classical limit of QMBS corresponds to classically unstable periodic orbits.  
   
In this paper, I will show a clear distinction between QMBS and classical chaos using the out-of-time ordered correlator 
(OTOC)~\cite{Larkin1969QuasiclassicalMI,Garc_a_Mata_2023} (the precise definition is given in Eq.\,\eqref{eq:OTOC in this paper} in Sec.\,\ref{sec:prep}). 
More precisely, I will compute OTOCs for a generic chaotic single-particle system and a simple many-body system hosting QMBS originating from an SM-like structure.  
The early time behavior of the OTOC of an unstable periodic orbit in the single-particle system exhibits conventional Lyapunov growth, 
\begin{equation}\label{eq:OTOC QS intro}
    OTOC(t)\sim\hbar^2\# e^{2\lambda_Lt},
\end{equation}
where $\lambda_L$ is the Lyapunov exponent (LE) calculated classically. Here, the small parameter of the theory is 
$\hbar$. It is particularly important that the exponent $\lambda_L$ is a quantity of order $O(1)=O(\hbar^0)$. In contrast, 
the OTOC of the periodic orbit originating from the scar states does not exhibit such rapid growth. The early time behavior is instead given by
\begin{equation}\label{eq:OTOC QMBS intro}
    OTOC(t)\sim\frac{\#}{N}e^{\lambda t/N},
\end{equation}
where $1/N$, the inverse of the number of flavor degrees of freedom, is the small parameter of the theory. The semiclassical limit corresponds to $N\rightarrow\infty$, and thus 
the LE vanishes in this limit. Note that, similar to the QS scenario, the periodic orbit corresponds to a periodic solution of the classical equation of motion in the large-$N$ limit. Therefore, 
Eq.\,\eqref{eq:OTOC QMBS intro} implies that the periodic orbit is destabilized solely by quantum fluctuation. The (classically) vanishing LE turns out to be a distinctive feature 
of QMBS in this model. The model further hosts another classically unstable periodic orbit that is, however, not associated with a QMBS. 
The OTOC of this orbit satisfies the conventional form like Eq.\,\eqref{eq:OTOC QS intro},
\begin{equation}\label{eq:OTOC nonQMBS intro}
    OTOC(t)\sim\frac{\#}{N}e^{2\lambda_L t},
\end{equation} 
where $\lambda_L$ is an $O(1)$ quantity calculated from the quadratic action. 
Eqs.\,(\ref{eq:OTOC QS intro}-\ref{eq:OTOC nonQMBS intro}) indicate clearly the different behavior of the two periodic orbits. 

Another important question in the study of QMBS is the stability of scarring under perturbations to the Hamiltonian. Many existing toy models feature finely tuned algebraic 
structures that facilitate scarring, yet the effects of perturbations that disrupt these structures remain less understood. Given that these models are inherently 
non-integrable and often strongly interacting, studies on the fate of QMBS under perturbations have primarily relied on 
numerical simulations of small system sizes or perturbative analysis~\cite{Khemani_2019,Lin2020_perturbation,Kolb_2023,lerose2023}. The model in 
this paper provides a novel insight beyond these common approaches: when a perturbation disrupts the SM structure of the 
Hamiltonian, the scar states cease to be exact eigenstates. The behavoir of the OTOC is modified differently depending on perturbation strength: 
when the coupling constant is smaller than a certain threshold, indicating a weak perturbation, it is modified as,  
\begin{equation}\label{eq:OTOC pert intro}
    OTOC(t)\sim\frac{\#}{N}C(t),
\end{equation}
where $C(t)$ is a sub-exponential function with an $O(1)$ value. Eq.\,\eqref{eq:OTOC pert intro} implies that although the weak scar-breaking perturbation destabilizes the periodic orbit, 
the conventional Lyapunov growth is still absent. On the other hand, when the perturbation is larger than the threshold, the OTOC shows conventional exponential growth like Eq.\,\eqref{eq:OTOC QS intro} 
or Eq.\,\eqref{eq:OTOC nonQMBS intro}.

The difference in behavior of the OTOCs originates from the SM structure. The monodromy matrix of an unstable periodic orbit typically has at least one eigenvalue with a positive real part, leading to 
the destabilization of the periodic orbit. In cases where the monodromy matrix has only pure imaginary eigenvalues (note that for Hamiltonian systems, the sum of 
the eigenvalues must be zero), linear stability analysis is inconclusive, necessitating consideration of higher-order terms. However, stability analysis beyond linearization is challenging and often 
requires a case-by-case approach. Remarkably, the SM structure guarantees that higher-order terms do not destabilize the periodic orbit in the semiclassical limit. Therefore only quantum 
fluctuations can destabilize the orbit, implying the LE of order $1/N$. 
Once the weak scar-breaking perturbation is added, the monodromy matrix still does not show unstable modes while higher-order terms affect the periodic orbit, eventually destabilizing it. 
This implies that OTOCs do grow, but their growth is subexponential. 

\begin{table}
    \centering
    \begin{tabular}{| c | c | c | c | c |}\hline
        System & Parameter &  $OTOC(t)$ & Mechanism & Diagram \\ \hline
        QS (Sec.\,\ref{sec:qs}) & $\hbar$ & $\sim e^{O(t)}$ & Quadratic &\includegraphics[width=1cm,valign=c]{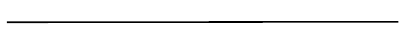}$\sim e^{\lambda t}$ \\ \hline
        QMBS (Sec.\,\ref{sec:QMBS}) & $1/N$ & $\sim e^{O(t/N)}$ & Second order & \includegraphics[width=1cm,valign=c]{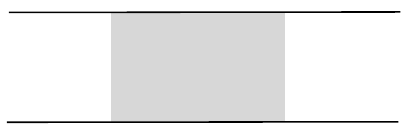}$\sim e^{\lambda t/N}$ \\ \hline
        Generic orbit (Sec.\,\ref{sec:unstable orbit}) & $1/N$ & $\sim e^{O(t)}$ & Quadratic & \includegraphics[width=1cm,valign=c]{bare_green.pdf}$\sim e^{\lambda t}$ \\ \hline
        Perturbed QMBS (Sec.\,\ref{sec:scar breaking perturbation}) & $1/N$ & subexponential & First order & \includegraphics[width=1cm,valign=c]{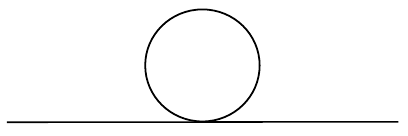}$\ll e^{\lambda t}$ \\ \hline        
    \end{tabular}    
    \caption{Summary of the different behaviors of the OTOCs and their origin.}
    \label{table:summary}
\end{table}

The rest of the paper is organized as follows. Sec.\,\ref{sec:prep} provides a concise overview of the basic concepts and technical tools 
utilized in this study. In Sec.\,\ref{sec:qs}, I calculate the OTOC for a generic single-particle system, under the
assumption of classical chaos, and demonstrate conventional Lyapunov growth of the OTOC,  
with the LE calculated obtained from classical calculations\,[Eq.\,\eqref{eq:OTOC QS intro}]. Sec.\,\ref{sec:model} introduces the many-body model, 
where I examine its thermodynamic properties and highlight an observable that indicates a violation of the ETH. 
The analysis of the scar states within this model is presented in Sec.\,\ref{sec:QMBS}, where the LE is derived via the Bethe-Salpeter (BS) equation, 
revealing its disappearance in the semiclassical limit. Sec.\,\ref{sec:unstable orbit} identifies another periodic orbit within the model, showing that 
the corresponding OTOC exhibits conventional Lyapunov growth. The impact of perturbation that breaks the scar structure is addressed in 
Sec.\,\ref{sec:scar breaking perturbation}. There I show that while the system exhibits chaotic dynamics even in the large-$N$ limit, conventional 
Lyapunov growth remains absent. The paper concludes with a summary and final remarks in Sec.\,\ref{sec:conclusion}.

\section{Preliminaries}\label{sec:prep}
\renewcommand{\theequation}{2.\arabic{equation}}
\setcounter{equation}{0}
\subsection{Shiraishi-Mori construction of many-body scars}
This paper focuses on the SM construction, one of the simplest methods for generating models that host QMBS. Other approaches, such as spectrum generating algebra (SGA)~\cite{eta_pair_rsga}, 
are reviewed in Ref.\,\cite{annurev-101617}. Consider a Hamiltonian of the form: 
\begin{equation}\label{eq:SM form}
    H=\sum_iP_ih_iP_i+H',
\end{equation}
where $P_i$ is a projector and $h_i$ is an arbitrary Hermitian operator. If there exists at least one state $\ket*{\psi}$ that is annihilated by all projectors $P_i$, i.e., 
$P_i\ket*{\psi}=0\,\,\forall i$, then the ``scar subspace'' $\mc{V}_{\mr{scar}}$ can be defined as the common kernel of $P_i$. If $\mc{V}_{\mr{scar}}$ is invariant under $H'$, 
the Hamiltonian $H$ will have $N_{\mr{scar}}=\dim\mc{V}_{\mr{scar}}$ eigenstate(s) $\ket*{S_n}$ within the scar subspace, i.e., $\ket*{S_n}\in\mc{V}_{\mr{scar}}\,(n=1\sim N_{\mr{scar}})$. 
The eigenstate(s) should satisfy the condition $\expval*{P_i}{S_n}=0$ by construction, violating the ETH since the thermal expectation value of $P_i$ at any non-zero temperature is 
strictly positive.

A natural extension of Eq.\,\eqref{eq:SM form} is to replace the projector with an operator that has a non-empty kernel. In this work, I set $P_i=b_i$, where 
$b_i$ is a bosonic annihilation operator. Consequently, the scar states satisfy the condition $\expval*{b^\dag_ib_i}{S_n}=0$. This extension, however, does not necessarily guarantee 
violation of the ETH. The infinite dimensionality of the local Hilbert space in bosonic systems and the unbounded nature (in terms of an operator norm) of local terms in the Hamiltonian 
invalidate the key technical ingredient of the proof of the ETH violation. To address this, I consider the large-$N$ limit, where the the model becomes classical 
and exactly solvable. In this regime, the expectation value of $b_i^\dag b_i$ can be directly calculated, demonstrating ETH vilation. Although this approach does not 
rigorously prove ETH violation for finite $N$, the thermal expectation value $\expval*{b_i^\dag b_i}$ should be close to its large-$N$ limit, with deviations of order $1/N$.

\subsection{Lyapunov exponent in out-of-time order correlator}
The OTOC was first introduced in the context of dirty superconductors in Ref.\,\cite{Larkin1969QuasiclassicalMI}, and later gained prominence for its role in studying quantum chaos, 
particularly in studying blackhole scrambling\,\cite{Shenker_2014}, and the conjectured chaos bound\,\cite{Maldacena_2016}. In classical mechanics,  
the LE $\lambda_L$ describes the exponential divergence of trajectories, captured by the Poisson bracket of observables at different times,
\begin{equation}\label{eq:BP OTOC}
    \{x(t),p(0)\}_{\mr{P.B.}}=\pdv{x(t)}{x(0)}\sim e^{\lambda_Lt},
\end{equation}
where $\{A,B\}_{\mr{P.B.}}$ is the Poisson bracket. In quantum mechanics, the Poisson bracket is replaced by the commutator ($\{,\}_{\mr{P.B.}}\mapsto(i\hbar)^{-1}[,]$), 
and the square of this operator defines the OTOC:
\begin{equation}
    OTOC(t)=\expval*{[x(t),p(0)][x(t),p(0)]^\dag},
\end{equation}
where $x(t)$ and $p(0)$ are now operators. The bracket represents the quantum mechanical average (see below for the precise definition used in this paper). 
From Eq.\,\eqref{eq:BP OTOC} one expects $OTOC(t)\sim\hbar^2e^{2\lambda_Lt}$. More generally, 
the OTOC can be defined for any pair of operators $V$ and $W$:
\begin{equation}
    OTOC(t)=\expval*{[V(t),W(0)][V(t),W(0)]^\dag}.
\end{equation} 

In the case of mixed states, there is some ambiguity in defining the expectation value. For instance, ``unregularized'' and ``regularized'' OTOCs are often 
distinguished as follows,
\begin{equation}
    \mr{Tr}\left(\rho[V(t),W(0)][V(t),W(0)]^\dag\right),\,\,\mr{Tr}\left(\sqrt{\rho}[V(t),W(0)]\sqrt{\rho}[V(t),W(0)]^\dag\right).
\end{equation}
These different definitions can lead to different prefactors in  
the OTOC~\cite{PhysRevB.98.205124}. Throughout this paper, however, all OTOCs are computed with respect to 
pure states. Specifically, I use the following definition of the OTOC relative to a reference state $\ket*{\psi}$,
 
\begin{equation}\label{eq:OTOC in this paper}
    OTOC(t)=\expval*{[V(t),W(0)][V(t),W(0)]^\dag}{\psi}.
\end{equation}

\section{Lyapunov growth in a single-particle semiclassical system}
\label{sec:qs}
\renewcommand{\theequation}{3.\arabic{equation}}
\setcounter{equation}{0}
In this section I study an OTOC of a single-particle semiclassical chaotic system and demonstrate that it exhibits conventional Lyapunov growth.
Quantum scarring refers to the concentration of eignfunctions on classically unstable periodic orbits in semiclassical billiards~\cite{Gutzwiller}. 
A single-particle wavepacket initialized on a weakly unstable periodic orbit exhibits anomalously long-lasting oscillations before eventually spreading 
out across the entire system~\cite{quantum_scar_original}.
These oscillations suggest the existence of scarred eigenstates. Following this approach, I consider a chaotic single-particle system described by the Hamiltonian 
$H=p^2/2m+U(x)$ where in the classical limit, the equation of motion admits an unstable periodic orbit with period $T$. Here the operators $x$ and $p$ satisfy the 
canonical commutation relation $[x,p]=i\hbar$ with $\hbar$ treated as a small parameter. Unlike the original setting of qunatum scarring, 
where $U(x)$ represents the sharp boundary of a billiard, 
I assume that the confining potential $U(x)$ is a smooth function of $x$. To compute the OTOC, I introduce the ladder operator,
\begin{equation}
    a=\sqrt{\frac{m\Omega}{2}}x+i\frac{p}{\sqrt{2m\Omega}},
\end{equation}
where $\Omega=2\pi/T$. This operator satisfies the commutation relation $[a,a^\dag]=\hbar$, and I denote the normal ordered Hamiltonian as
\begin{equation}
    H=\sum_{n=0}^\infty\hbar^n H_n(a^\dag,a).
\end{equation}
$H_n (n\geq1)$ is obtained by normal ordering the original Hamiltonian, 
and thus in the semiclassical limit ($\hbar\rightarrow0$) it should vanish. The classical equation of motion is recovered by 
neglecting correlations (e.g., $\expval*{a^n}\approx\expval{a}^n$) of $H_0$. The equation for $\expval*{a}$ 
within this approximation is written as
\begin{equation}\label{eq:classical eom}
    i\dv{\expval*{a}}{t}\approx\pdv{H_0(\expval*{a^\dag},\expval*{a})}{\expval*{a^\dag}},
\end{equation}
which is equivalent to the classical equation of motion, or the Ehrenfest theorem. By assumption Eq.\,\eqref{eq:classical eom} 
has a periodic solution, which I denote $\alpha_0(t)$. The stability of this solution can be inspected by checking the eigenvalues 
of the monodromy matrix $M(t)$, which is the Hessian of $H_0$ with respect to the periodic solution. Namely, $M(t)$ is defined as
\begin{equation}\label{eq:monodromy}
    M(t)=\left.\begin{pmatrix}
        \pdv{H_0}{\expval*{a^\dag}}{\expval*{a}}&\pdv{H_0}{\expval*{a^\dag}}{\expval*{a^\dag}}\\ \pdv{H_0}{\expval*{a}}{\expval*{a}} & \pdv{H_0}{\expval*{a}}{\expval*{a^\dag}}
    \end{pmatrix}\right|_{(\expval*{a},\expval*{a^\dag})=(\alpha_0(t),\alpha^*_0(t))}.
\end{equation}

\begin{figure}
    \centering
    \includegraphics*[width=0.7\linewidth]{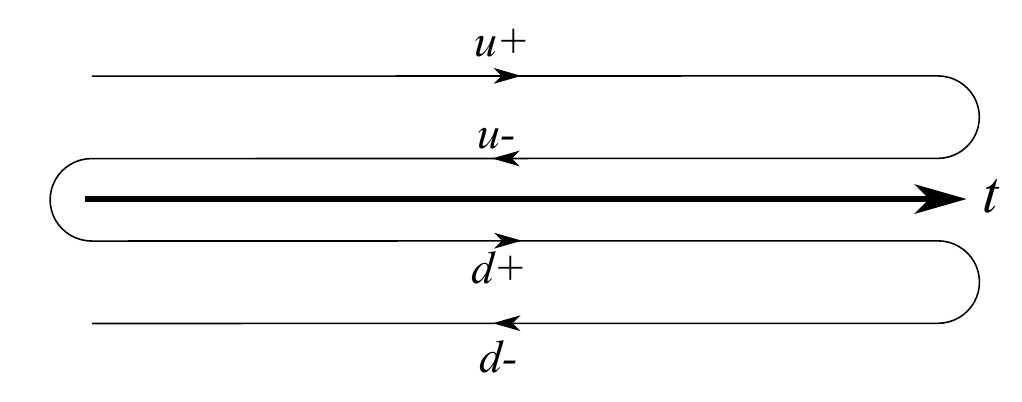}
    \caption{Two folded Keldysh contour. The arrows indicate the path-ordering of operators 
    in the Keldysh formalism.}
    \label{fig:keldysh path}
\end{figure}

I will compute the following OTOC $C(t)$,
\begin{equation}\label{eq:OTOC QS}
    C(t)=\theta(t)\expval*{[a(t),a^\dag(0)][a(t),a^\dag(0)]^\dag}{g},
\end{equation}
where $\ket*{g}$ is a Gaussian wavepacket centered on the periodic orbit. 
In the following I will sketch the important steps of the derivation, and leave details in Appendix\,\ref{app:OTOC QS}. 
$C(t)$ can be computed perturbatively with the small parameter $\hbar$. 
Technically this perturbative expansion should be carried out in the two-folded Keldysh contour 
(Fig.\,\ref{fig:keldysh path}), as Eq.\,\eqref{eq:OTOC QS} is not time-ordered. However this is not necessary in order 
to obtain the LE only to the leading order in $\hbar$ as shown below. 

\begin{figure}
    \centering
    \includegraphics[width=\linewidth]{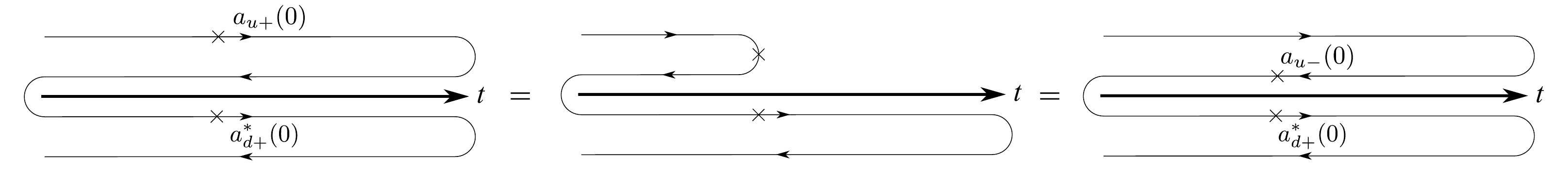}
    \caption{A graphical proof of the equality $\expval*{a_{d+}^*(0)a_{u-}(0)}=\expval*{a^*_{d+}(0)a_{u+}(0)}$ (the first line of Eq.\,\eqref{eq:equality}). 
    Due to the unitarity, part of the $(u+)$- and $(u-)$-branches can be cancelled out, allowing the Keldysh contour to be folded 
    immediately after time $0$.}
    \label{fig:two-folded Keldysh equality}
\end{figure}
To compute Eq.\,\eqref{eq:OTOC QS} I define the retarded Green's function $G^R$,
\begin{equation}\label{eq:retarded QS}
    iG^R(t,0)=\theta(t)\expval*{[a(t),a^\dag(0)]}{g}.
\end{equation}
$G^R$ can be computed perturbatively with respect to $\hbar$: in this approach the unperturbed part of the Hamiltonian 
corresponds to the monodromy matrix $M(t)$ and all the other terms are regarded as perturbations.
Note that $G^R$ directly enters the OTOC: for example, at the zeroth order of Eq.\,\eqref{eq:OTOC QS} is a product of $G^R$,
\begin{equation}\label{eq:bare OTOC qs}
    C^{(0)}(t)=G^R_0(t,0)(G^R_0(t,0))^*,
\end{equation}
where $G^R_0(t,0)$ is the bare Green's function. Since this form will appear frequently in subsequent OTOC computations, 
here I provide a detailed derivation: the OTOC is decomposed as
\begin{equation}\begin{split}\label{eq:OTOC decomp}
    C(t)&=\theta(t)\big(\expval*{a(t)a^\dag(0)a(0)a^\dag(t)}{g}-\expval*{a(t)a^\dag(0)a^\dag(t)a(0)}{g}\big)\\
    &-\theta(t)\big(\expval*{a^\dag(0)a(t)a(0)a^\dag(t)}{g}-\expval*{a^\dag(0)a(t)a^\dag(t)a(0)}{g}\big).
\end{split}
\end{equation}
These four-point functions can be written as products of two-point functions by Wick theorem. 
For instance, within the zeroth order in the perturbation theory, the first two terms in Eq.\,\eqref{eq:OTOC decomp} 
in the path-integral formalism are written as
\begin{equation}\label{eq:Wick decomp}\begin{split}
    \expval*{a(t)a^\dag(0) a(0)a^\dag(t)}{g}&=\expval*{a_{d-}(t)a^*_{d+}(0)a_{u-}(0)a^*_{u+}(t)}\\
    &=\expval*{a_{d-}(t)a_{d+}^*(0)}\expval*{a_{u-}(0)a^*_{u+}(t)}
    +\expval*{a_{d-}(t)a^*_{u+}(t)}\expval*{a^*_{d+}(0)a_{u-}(0)}\\
    &+\expval*{a_{d-}(t)a_{u-}(0)}\expval*{a^*_{d+}(0)a^*_{u+}(t)}\\
    \expval*{a(t)a^\dag(0)a^\dag(t)a(0)}{g}&=\expval*{a_{d-}(t)a^*_{d+}(0)a^*_{u-}(t)a_{u+}(0)}\\
    &=\expval*{a_{d-}(t)a^*_{d+}(0)}\expval*{a^*_{u-}(t)a_{u+}(0)}
    +\expval*{a_{d-}(t)a_{u+}(0)}\expval*{a^*_{d+}(0)a^*_{u-}(t)}\\
    &+\expval*{a_{d-}(t)a^*_{u+}(t)}\expval*{a_{d+}^*(0)a_{u+}(0)},
\end{split}
\end{equation}
where the right-hand side (RHS) represents the path-integral formulation with 
the bosonic fields $a_{\gamma},a^*_\gamma\,(\gamma=(u\pm),(d\pm))$ residing in the $\gamma$-branch 
of the Keldysh contour. The unitarity of the time-evolution permits deformations of the Keldysh contour, 
leading to the following identities (see Fig.\,\ref{fig:two-folded Keldysh equality} for the graphical 
representation),
\begin{equation}\label{eq:equality}\begin{split}
    \expval*{a_{d+}^*(0)a_{u-}(0)}&=\expval*{a^*_{d+}(0)a_{u+}(0)}\\
    \expval*{a_{d-}(t)a_{u-}(0)}&=\expval*{a_{d-}(t)a_{u+}(0)}\\
    \expval*{a_{d+}^*(0)a_{u+}^*(t)}&=\expval*{a^*_{d+}(0)a_{u-}^*(t)}.
\end{split}
\end{equation} 
Substituting these equalities into Eq.\,\eqref{eq:Wick decomp}, the first line of the RHS 
in Eq.\,\eqref{eq:OTOC decomp} takes the following simple form,
\begin{equation}\begin{split}
    &\theta(t)\big(\expval*{a(t)a^\dag(0)a(0)a^\dag(t)}{g}-\expval*{a(t)a^\dag(0)a^\dag(t)a(0)}{g}\big)\\
    &=\theta(t)\expval*{a_{d-}(t)a^*_{u+}(0)}\expval*{\big(a_{u-}(0)a_{u+}^*(t)-a^*_{u-}(t)a_{u+}(0)\big)}\\
    &=\theta(t)\expval*{a_{d-}(t)a^*_{u+}(0)}(\expval*{[a(t),a^\dag(0)]}{g})^*\\
    &=-i\theta(t)\expval*{a_{d-}(t)a^*_{d+}(t)}(G_0^R(t,0))^*.
\end{split}
\end{equation}  
The same procedure can be applied to the second line of Eq.\,\eqref{eq:OTOC decomp}, yielding 
Eq.\,\eqref{eq:bare OTOC qs}.

As shown in Appendix\,\ref{app:OTOC QS}, calculation of the bare Green's function is equivalent to 
finding the eigenmodes of the monodromy matrix,
\begin{equation}\label{eq:classical monodromy}
    i\sigma^3\dv{\psi_\alpha(t)}{t}=M(t)\psi_\alpha(t),
\end{equation}
where $\sigma^3=\mr{diag}(1,-1)$ is the Pauli matrix.
Since $M(t)$ is periodic in time, the Floquet theorem applies and $\psi_\alpha(t)$ is written as 
$\psi_\alpha(t)=e^{\lambda_\alpha t}u_\alpha(t)$ where $u_\alpha(t)$ is a periodic function. The 
exponential factor with the largest real part ($\max_\alpha\Re\lambda_\alpha$) corresponds to the (classical) LE of the periodic 
orbit. Note that Eq.\,\eqref{eq:bare OTOC qs} already shows the exponential growth: using the normal modes $\psi_\alpha(t)$, 
the OTOC is written as,
\begin{equation}\label{eq:OTOC lyapunov qs}
    C^{(0)}(t)=\hbar^2\theta(t)\left|\sum_\alpha e^{\lambda_\alpha t}\left[u_\alpha(t)u^\dag_\alpha(0)\right]_{11}\right|^2,
\end{equation}
where $[A]_{11}$ indicates the $(1,1)$-component of the matrix $A$. 

Eq.\,\eqref{eq:OTOC lyapunov qs} and Eq.\,\eqref{eq:classical monodromy} show that, to the leading order (in $\hbar$), the OTOC is 
equivalent to the classical problem, i.e., finding the eigenmodes of the monodromy matrix.
I emphasize that in this case the Lyapunov exponent $\Re\lambda_\alpha$ is $O(1)$, which is not affected by the semiclassical limit ($\hbar\rightarrow0$).
For QMBS the situation is rather different. In Sec.\,\ref{sec:QMBS}, I will show that the OTOC for QMBS at the 
leading order does not show any exponential divergence, and only quantum corrections are 
responsible for the Lyapunov growth of the OTOC, which implies that in the semiclassical limit they vanish.  

\section{Many-body model and thermodynamics}
\label{sec:model}
\renewcommand{\theequation}{4.\arabic{equation}}
\setcounter{equation}{0}

I consider a one-dimensional $N$-flavor bosonic system with periodic boundary conditions,
\begin{equation}\label{eq:H}
    \begin{split}
        H&=H_0+V\\
        H_0&=-t\sum_{i=1}^L\sum_{\mu=1}^N\left(a_{i\mu}^\dag a_{i+1\mu}+a_{i+1\mu}^\dag a_{i\mu}
        +b_{i\mu}^\dag b_{i+1\mu}+b^\dag_{i+1\mu}b_{i\mu}\right)
        +\mu\sum_{i=1}^L\sum_{\mu=1}^N\left(a_{i\mu}^\dag a_{i\mu}+b^\dag_{i\mu}b_{i\mu}\right)\\
        &\equiv\sum_{k\in\mr{BZ}}\sum_{\mu=1}^N\xi_k\left(a_{k\mu}^\dag a_{k\mu}+b_{k\mu}^\dag b_{k\mu}\right)\\
        V&=\sum_{i=1}^Lv_i,
    \end{split}
\end{equation}
where $a_{i\mu}(a_{k\mu})$ and $b_{i\mu}(b_{k\mu})$ are bosonic annihilation operators at the site $i$ 
(in momentum space) and $\xi_k=-2t\cos k+\mu>0$ is a dispersion. The local potential, $v_i$, is defined as
\begin{equation}\label{eq:interaction}
    v_i=\frac{J}{N^2}\sum_{\mu,\nu,\rho=1}^N
    \normord{b_{i\mu}^\dag b_{i\mu}(a_{i\nu}+a_{i\nu}^\dag)^2(b_{i\rho}+b_{i\rho}^\dag)^2},
\end{equation}
where $\normord{\bullet}$ is the normal order and $J$ is set positive. The prefactor $J/N^2$ 
is chosen so that the potential energy is proportional to $N$ ($\expval*{v_i}=O(N)$). This model has a parameter $N$, 
the number of flavors, and in the large-$N$ limit the saddle-point (or the mean-field) approximation becomes exact. 
Therefore $1/N$ plays the role of $\hbar$ in QS, and for this reason I set $k_B=\hbar=1$. 

This model hosts exponentially many scar states, as shown in Sec.\,\ref{sec:QMBS}. They 
consist solely of $a$-bosons and thus annihilated by each local potential $v_i$ since $v_i$ must 
contain at least one annihilation operator of $b$-bosons. The absence of $b$-bosons implies 
that the scar states are not thermal, as discussed below.

Before proceeding with the analysis of this model, a few preliminary remarks are necessary. 
The specific form of the interaction $v_i$ does not appear to be crucial; 
different potentials should yield similar OTOC behavior to those summarized in 
Table\,\ref{table:summary}, as long as each term $v_i$ annihilates the scar states  
defined in Eq.\,\eqref{eq:scar subspace} in Sec.\,\ref{sec:QMBS}, i.e., the SM structure described in Sec.\,\ref{sec:QMBS} is preserved. 
However, it is preferable to choose a potential that is lower-bounded, 
and explicitly breaks the $U(1)$ symmetry of $a$- and $b$-bosons. For bosonic models, lower-bounded 
potentials are generally required so that the free energy is lower-bounded. Since the OTOCs defined in Eq.\,\eqref{eq:otoc QMBS} and 
Eq.\,\eqref{eq:OTOC another orbit} below correspond to perturbations of periodic orbits by adding a single boson, maintaining $U(1)$ symmetry could significantly constrain the phase space  
the system can explore, potentially suppressing chaos (see also the discussion following Eq.\,\eqref{eq:otoc QMBS} in Sec.\,\ref{sec:QMBS}). 
As discussed below, the model can be considered a simple deformation in that the potential $v_i$ only breaks the system's 
$U(1)^{2N}$ symmetry of the non-interacting part.

\subsection{Spontaneous symmetry breaking}
While this model does not possess the full $U(1)$ symmetry of $a$- and $b$-bosons, it has an $O(N)^2$ symmetry, 
which is represented by the following action,
\begin{equation}\begin{split}
    \frac{a_{i\mu}+a^\dag_{i\mu}}{2}\rightarrow\sum_{\nu=1}^NU_{\mu\nu}\frac{a_{i\nu}+a^\dag_{i\nu}}{2},\,\,
    \frac{a_{i\mu}-a^\dag_{i\mu}}{2i}\rightarrow\sum_{\nu=1}^NU_{\mu\nu}\frac{a_{i\nu}-a^\dag_{i\nu}}{2i}\\
    \frac{b_{i\mu}+b_{i\mu}^\dag}{2}\rightarrow\sum_{\nu=1}^NV_{\mu\nu}\frac{b_{i\nu}+b^\dag_{i\nu}}{2},\,\,
    \frac{b_{i\mu}-b^\dag_{i\mu}}{2i}\rightarrow\sum_{\nu=1}^NV_{\mu\nu}\frac{b_{i\nu}-b^\dag_{i\nu}}{2i},
\end{split}
\end{equation}
where $U$ and $V$ are distinct elements of $O(N)$. 
The symmetry of the zero-temperature state is studied by the Euler-Langrange (EL) equation 
for $a_\mu=\expval*{a_{i\mu}},b_\mu=\expval*{b_{i\mu}}$, which reads,
\begin{equation}\label{eq:EL eq}
    \begin{split}
        &E a_\mu+2J\left(\frac{1}{N}\sum_{\nu=1}^N|b_\nu|^2\right)\left(\frac{1}{N}\sum_{\rho=1}^N(b_\rho+b^*_\rho)^2\right)(a_\mu+a^*_\mu)=0\\
        &E b_\mu+J\left(\frac{1}{N}\sum_{\nu=1}^N(a_\nu+a^*_\nu)^2\right)
        \left[\left(\frac{1}{N}\sum_{\rho=1}^N(b_\rho+b^*_\rho)^2\right)b_\mu+\left(\frac{2}{N}\sum_{\rho=1}^N|b_\rho|^2\right)(b_\mu+b^*_\mu)\right]=0,
    \end{split}
\end{equation}
where $E=-2t+\mu$ is the dispersion at $k=0$. The first line implies $a_\mu\in\mb{R}$, as the second term is real so that the equation is written as $Aa_\mu=B$ 
with $A,B\in\mb{R}$. This in turn implies $b_\mu\in\mb{R}$ by a similar reasoning. The reality of $a_\mu$ and $b_\mu$ simplifies Eq.\,\eqref{eq:EL eq} further 
as
\begin{equation}\label{eq:EL reality}
    \begin{split}
        &\left(E+16J\left(\frac{1}{N}\sum_{\nu=1}^Nb_\nu^2\right)^2\right)a_\mu=0\\
        &\left(E+32J\left(\frac{1}{N}\sum_{\nu=1}^Na_\nu^2\right)\left(\frac{1}{N}
        \sum_{\rho=1}^Nb_\rho^2\right)\right)b_\mu=0.
    \end{split}
\end{equation}
When $E>0$, Eq.\,\eqref{eq:EL reality} has the unique solution $a_\mu=b_\mu=0$. When $E<0$, 
which will however not be relevant for the subsequent 
discussion, the solution of Eq.\,\eqref{eq:EL reality} is 
\begin{equation}
    \frac{1}{N}\sum_{\mu=1}^Na_\mu^2=\frac{1}{8}\sqrt{\frac{-E}{J}},\,\,
    \frac{1}{N}\sum_{\mu=1}^Nb_\mu^2=\frac{1}{4}\sqrt{\frac{-E}{J}},
\end{equation} 
which breaks the $O(N)^2$ symmetry. 

\subsection{Mean-field analysis: equilibrium}
Since the Euclidean action of the model $S_E$ is proportional to $N$, the saddle-point 
(or mean-field) approximation becomes exact in the large-$N$ limit. In this limit, the thermal 
expectaion value of an observable $O$ is simply given by the saddle-point value,
\begin{equation}\begin{split}
    \lim_{N\to\infty}\expval*{O}
    &\equiv\lim_{N\to\infty}\int D[a^*,a]D[b^*,b]D\Delta_a Dv_a D\Delta_b Dv_b Oe^{-S_E}
    \\&=\lim_{N\to\infty}\int D[a^*,a]D[b^*,b]Oe^{-S_E^*},
\end{split}
\end{equation}
where $\Delta_{a/b}$ and $v_{a/b}$ are auxiliary fields that make the action quadratic in 
$a$ and $b$ (see Appendix\,\ref{app:thermal derivation} for the detailed discussion), 
and $S_E^*$ is obtained by substituting the saddle-point value of the auxiliary fields 
into $S_E$.

Within the mean-field approximation, one can show that the density of the $b$-bosons 
(see Appendix\,\ref{app:thermal derivation}),
\begin{equation}\label{eq:b boson density}
    \rho_b=\frac{1}{NL}\sum_{i=1}^L\sum_{\mu=1}^Nb_{i\mu}^\dag b_{i\mu},
\end{equation}
has to be strictly positive for any $T$, i.e., $\expval{\rho_b}_T>0$. In Sec.\,\ref{sec:QMBS} below I will define 
the scar states [Eq.\,\eqref{eq:scar subspace}], and show that they are characterized by the property 
$\rho_b\ket*{\psi}=0$, meaning that the scar states cannot be thermal eigenstates at any $T$. 

Strictly speaking, the treatment in this section is insufficient to prove violation of the ETH in a rigorous manner: 
the positivity of $\rho_b$ at any temperature can be shown only for the large-$N$ limit, while OTOCs are computed for large but 
finite $N$. However, it is reasonable to assume that the difference between the finite $N$ and the large-$N$ results is subleading in $N$ 
and thus $\expval*{\rho_b}_T$ should still be strictly positive. 

\section{Scar states and Lyapunov exponent}
\label{sec:QMBS}
\renewcommand{\theequation}{5.\arabic{equation}}
\setcounter{equation}{0}

Any state containing solely $a$-bosons is annihilated by each local potential $v_i$ as $b_i\in v_i$. This type of construction of the scarred 
Hamiltonian can be viewed as a variant of the SM construction since the scar subspace is a common kernel of each local operator $v_i$. 
In the present case, the following Fock space $\mc{F}_a$ is the scar subspace,
\begin{equation}\label{eq:scar subspace}
    \mc{F}_a=\mr{span}\left\{\prod_{k\in\mr{BZ}}\prod_{\mu=1}^N\big(a^\dag_{k\mu}\big)^{n_{k\mu}}\ket*{0}\bigg|\,n_{k\mu}\in\mb{Z}_{\geq0}\right\},
\end{equation}
where $\ket*{0}$ is the vacuum state. Within $\mc{F}_a$, only the non-interacting part of the Hamiltonian acts non-trivially. As a result, the 
Fock states of $a$-bosons in the momentum basis are exact eigenstates. This structure is common in many toy models of QMBS, where 
the Hamiltonian is typically a sum of a Zeeman term and locally annihilating terms. Any state $\ket*{\psi}$ in $\mc{F}_a$ violates ETH since 
$\expval*{\rho_b}{\psi}=0$ while the thermal expectation value is strictly positive at any temperature. 
A simple consequence of the existence of the scar states is the failure of thermalization when the system is initialized within the scar 
subspace, i.e., a state containing only the $a$-boson. 
Specifically let us consider the following coherent state $\ket*{\alpha}$ as the initial state, 
\begin{equation}\label{eq:scar initial state}
    a_{i\mu}\ket*{\alpha}=\alpha_\mu\ket*{\alpha},\,\,b_{i\mu}\ket*{\alpha}=0\,\,\,\,\forall i,\mu.
\end{equation}
This state will never thermalize. The time-evolution is given by a simple precession of the coherent state,
\begin{equation}\label{eq:scar evolution}
    \ket*{\alpha(t)}=\ket*{\alpha e^{-iEt}},\,\,E=-2t+\mu\,\,(\forall N).
\end{equation}
Indeed, the Heisenberg equations projected onto $\mc{F}_a$ read,
\begin{equation}
    i\dv{\expval*{a_{i\mu}(t)}{\alpha}}{t}=E\expval*{a_{i\mu}(t)}{\alpha},\,i\dv{b_{i\mu}(t)}{t}=0,
\end{equation} 
without any approximation. In the Keldysh path-integral formalism, the solution 
($a_{i\mu}(t)=e^{-iEt}\alpha_\mu$ and $b_{i\mu}(t)=0$) satisfies the EL euqation of the Keldysh action 
(see Eq.\,\eqref{eq:scar solution} in Appendix\,\ref{app:Lyapunov QMBS}).
Note that the energy of this state is conserved and expressed as,
\begin{equation}
    \frac{\expval*{H}{\alpha(t)}}{NL}=\frac{E}{N}\sum_{\mu=1}^N|\alpha_\mu|^2\equiv E\alpha^2.
\end{equation}
Recall that $E=-2t+\mu>0$. The energy of the state $\ket*{\alpha(t)}$ corresponds to that of 
a high temperature state if $\alpha^2$ is sufficiently large, as I will assume in what follows. 

To study quantum chaos around the scar state $\ket*{\alpha(t)}$, I will compute the following OTOC $C_\alpha(t)$,
\begin{equation}\label{eq:otoc QMBS}
    C_\alpha(t)=\frac{\theta(t)}{N^2}\sum_{i=1}^L\sum_{\mu,\nu=1}^N\expval*{[b_{i\mu}(t),b_{j\nu}^\dag(0)][b_{i\mu}(t),b_{j\nu}^\dag(0)]^\dag}{\alpha}.
\end{equation}
The choice of the operators for OTOC is motivated by the following observation: let $C_{VW}(t)$ be the OTOC with respect to the operators $V$ and $W$,
\begin{equation}
    C_{VW}(t)=\frac{\theta(t)}{N^2}\sum_{i=1}^L\sum_{\mu,\nu=1}^N\expval*{[V(t),W(0)][V(t),W(0)]^\dag}{\alpha},
\end{equation}
where the above choice [Eq.\,\eqref{eq:otoc QMBS}] corresponds to $V=b,W=b^\dag$. One can immediately find that the choice $V=a,W=a^\dag$ 
(or $V=a^\dag,W=a$) does not show any chaos, as the dynamics is closed within the scar subspace. 
Another choice $V=a, W=b^\dag$ does not reveal any information of the system either, since
\begin{equation}
    C_{a,b^\dag}(t)=\frac{\theta(t)}{N^2}\sum_{i=1}^L\sum_{\mu,\nu=1}^L\expval*{a(t)b^\dag ba^\dag(t)}{\alpha}=0,
\end{equation} 
due to the property of the scar states $b\ket*{\psi}=0$. While the choice $V=b,W=a^\dag$ does lead to exponential growth of the OTOC, 
the exponent is smaller than the one in $C_\alpha(t)$ when $\alpha$ is large enough. Therefore, I will compute the particular choice of the OTOC, 
i.e., $C_\alpha(t)$ as it is expected to exhibit the largest LE.

The $U(1)$-breaking interaction [Eq.\,\eqref{eq:interaction}] is expected to favor chaoticity by lifting symmetry-related restrictions. The dynamics of the system would 
be otherwise restricted, as the reference state $\ket*{\alpha}$ does not contain any $b$-bosons. Specifically, the dimension of the Fock space 
of one-particle $b$-boson states $\mc{F}_b^1$, which would constitute the accessible part of the Hilbert space to the system if $U(1)$ were preserved, 
is $NL$ ($\dim\mc{F}^1_b=NL$). This is significantly smaller than the typical dimension of the many-body Hilbert 
space within the energy shell corresponding to a given finite temperature, which scales as $e^{O(NL)}$. In the following I will demonstrate that, 
even without such entropic protection, QMBS exhibit considerable stability due to the underlying SM structure.

Following equilibrium cases\,\cite{Stanford_2016,Patel_2017}, $C_\alpha(t)$ is calculated by the perturbation theory in the interaction $J$ in the two-folded Keldysh contour (Fig.\,\ref{fig:keldysh path}). Since the calculation is lengthy 
but not worth inspecting except for Eq.\,\eqref{eq:full RGF} and Eq.\,\eqref{eq:BS equation main text}, I will present only the important steps and results, 
and leave details in Appendix.\,\ref{app:Lyapunov QMBS}.

\subsection{Retarded Green's function}
I first calculate the retarded Green's function $G^R$ up to the second order in $J$. $G^R$ is defined as 
\begin{equation}
    iG^R(k;t,t')=\theta(t-t')\expval*{[b_{k\mu}(t),b^\dag_{k\mu}(t')]}{\alpha}.
\end{equation}
Note that due to the permutation symmetry of the Hamiltonian, one can omit the flavor index $\mu$. As in Sec.\,\ref{sec:qs}, $G^R$ directly enters the expression of the OTOC: for example, at the zeroth order of $J$, it is written as,
\begin{equation}\label{eq:OTOC zero qmbs}
    C_S^{(0)}(t)=\frac{1}{NL}\sum_{k\in\mr{BZ}}G^R_0(k;t,0)(G^R_0(k;t,0))^*,
\end{equation}
where $G^R_0$ is the bare Green's function. One can derive Eq.\,\eqref{eq:OTOC zero qmbs} 
in a similar manner to Eq.\,\eqref{eq:bare OTOC qs}.

\subsubsection{The first order correction}\label{subsec:1st order}
\begin{figure}
    \begin{minipage}{0.48\linewidth}
            \includegraphics*[width=\linewidth]{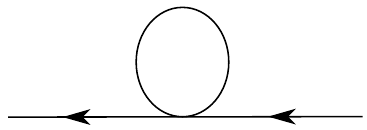}
    \end{minipage}
    \begin{minipage}{0.48\linewidth}
        \includegraphics[width=\linewidth]{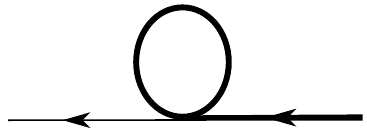}
    \end{minipage}
    \caption{The Feynman diagram of the first order perturbation (left) and the Hartree diagram (right). Only these contributions will survive in the large-$N$ limit, 
    However, the both diagrams vanish for the scar state. Once the scar-breaking perturbation is added, these diagrams affect the dynamics (see Sec.\,\ref{sec:scar breaking perturbation}).}
    \label{fig:1st order}
    \vspace*{1cm}
    \includegraphics[width=\linewidth]{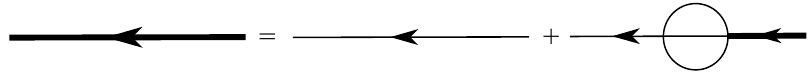}
    \caption{A graphical representation of the Dyson equation. Since the Hartree diagram should always vanish due the property of the scar state $b_{i\mu}(t)\ket*{\alpha}=0$,
    it consists only of second order perturbations (up to $1/N$).}
    \label{fig:Hartree}
\end{figure}

The first order perturbation corresponds to the diagram shown in the left panel of Fig.\,\ref{fig:1st order}. 
Furthermore, when all the corresponding higher corrections are taken into account, one arrives at the Hartree diagram, the right panel of Fig.\,\ref{fig:1st order}. This correction usually yields the mean-field equation. 
However, it is important that these corrections vanish for the scar states. 
As one can read from the diagram, the corrections contain normal ordered quadratic terms $\expval*{b^\dag_k(t)b_k(t)},\expval*{b_{-k}(t)b_k(t)}$, and $\expval*{b^\dag_k(t)b_{-k}^\dag(t)}$, 
which are always zero since $b_k$ annihilates the scar states. 

The vanishing the Hartree term turns out to be the most important characteristics of the (exact) scar states: 
as in the single-particle case in Sec.\,\ref{sec:qs}, bare Green's functions exhibit exponential divergence 
for generic unstable periodic orbits. This is also true for the present model, as discussed in Sec.\,\ref{sec:unstable orbit}. 
In rare cases where bare Green's functions do not diverge exponentially, \textit{full} Green's functions 
are still expected to diverge, although such divergence may be subexponential. The absence of exponential divergence in bare Green's functions  
and subexponential growth in full Green's functions are indeed observed when a weak perturbation is added, as discussed in Sec.\,\ref{sec:scar breaking perturbation}. 
On the other hand, the amplitude of even the full Green's function of the exact QMBS remains constant in the semiclassical limit ($N\to\infty$). 
Moreover, as shown below, the absence of growth in the Green's function persists even when quantum corrections (up to order $1/N$) are taken into account. 
\subsubsection{The second order correction}
\begin{figure}
    \begin{minipage}{0.32\linewidth}
        \includegraphics[width=\linewidth]{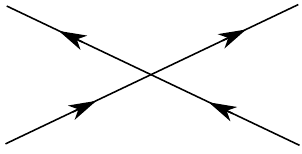}
    \end{minipage}
    \begin{minipage}{0.32\linewidth}
        \includegraphics[width=\linewidth]{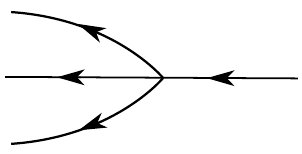}
    \end{minipage}
    \begin{minipage}{0.32\linewidth}
        \includegraphics[width=\linewidth]{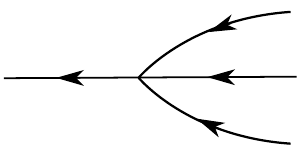}
    \end{minipage}
    \caption{Three scattering processes relevant for the second order corrections to 
    the Green's function: the number conserving two-body scattering (left), the number increasing 
    scattering (middle), and the number decreasing scattering (right).}
    \label{fig:3 processes}
\end{figure}
The second order corrections correspond to the self-energy that is order $1/N$.  
The Dyson equation is
\begin{equation}
    G^R(t,t')=G^R_0(t,t')-\int_{-\infty}^\infty dt_1dt_2G^R_0(t,t_1)\Sigma(t_1,t_2)G^R(t_2,t').
\end{equation}
Under the assumption that $\alpha^2$ is sufficiently large, the dominant contribution to $\Sigma$ is written as, 
\begin{equation}\label{eq:self-energy}\begin{split}
    \Sigma(t_1,t_2)&\approx\frac{8J^2\alpha^4}{N}\cos^2(Et_1)\cos^2(Et_2)
    \tilde{\Sigma}[G_0],     
\end{split}
\end{equation}
where $\tilde{\Sigma}[G_0]$ is a functional of the bare Green's function defined in Eq.\,\eqref{eq:Sigma functional} in Appendix\,\ref{app:Lyapunov QMBS}. 
Note that for the early time Lyapunov growth of the OTOC, it is sufficient to consider the self-energy consisting of the bare Green's functions $\tilde{\Sigma}[G_0]$, 
instead of the full Green's function ($\tilde{\Sigma}[G]$). In the momentum space the Dyson equation becomes 
\begin{equation}\label{eq:Dyson momentum}
    G^R(k,\omega)=G^R_0(k,\omega)-G_0^R(k,\omega)\sum_{n\in\mb{Z}}\Sigma_n(k,\omega)G^R(k,\omega+2nE),
\end{equation}
which resembles the Schr\"odinger equation of periodically driven systems. Indeed, the $2nE$ shift of the frequency comes from 
the cosines, the effect of the periodic orbit. While the exact solution of Eq.\,\eqref{eq:Dyson momentum} seems formidable, a perturbative calculation is still possible, and 
it is sufficient to determine the LE up to the order of $1/N$. 

Eq.\,\eqref{eq:Dyson momentum} can be written in vector form,
\begin{equation}
    \begin{pmatrix}
        \vdots\\ G^R(k,\omega-2E)\\ G^R(k,\omega) \\ G^R(k,\omega+2E)\\ \vdots
    \end{pmatrix}
    =\begin{pmatrix}
        \vdots\\ G_0^R(k,\omega-2E)\\ G^R(k,\omega)\\ G^R(k,\omega+2E)\\ \vdots 
    \end{pmatrix}
    -\msf{G}_0^R(k)\msf{\Sigma}(k)\begin{pmatrix}
        \vdots\\ G^R(k,\omega-2E)\\ G^R(k,\omega) \\ G^R(k,\omega+2E)\\ \vdots 
    \end{pmatrix},
\end{equation}
where 
\begin{equation}
    \begin{split}
        \msf{G}_0^R(k)&=\mr{diag}(\cdots,G_0^R(k,\omega-2E),G_0^R(k,\omega),G_0^R(k,\omega+2E),\cdots)\\
        \msf{\Sigma}(k)&=\begin{pmatrix}
      \ddots&                     &                  &                     & \\
      \cdots&\Sigma_{-1}(k,\omega)&\Sigma_0(k,\omega)&\Sigma_{+1}(k,\omega)&\cdots \\
            &            \cdots   &\Sigma_{-1}(k,\omega+2E)&\Sigma_0(k,\omega+2E)&\Sigma_{+1}(k,\omega+2E)&\cdots \\
            &                     &                        &                     &                        &\ddots
        \end{pmatrix}.
    \end{split}
\end{equation}
I decompose the ``matrix'' $\msf{\Sigma}$ as the diagonal part $\msf{D}_\Sigma$ and the non-diagonal part $\msf{N}_\Sigma$, i.e., $\msf{\Sigma}=\msf{D}_\Sigma+\msf{N}_\Sigma$.
Eq.\,\eqref{eq:Dyson momentum} is formally solved by inverting the matrix $1+\msf{G}_0^R\msf{\Sigma}$, and the effect of the non-diagonal part can be incorporated perturbatively, 
\begin{equation}\begin{split}\label{eq:GS dyson}
    \left(1+\msf{G}_0^R(k)\msf{\Sigma}(k)\right)^{-1}&=(1+\msf{G}_0^R(k)\msf{D}_\Sigma(k))^{-1}\\
    &+\sum_{n=1}^\infty\left((1+\msf{G}_0^R(k)\msf{D}_\Sigma(k))^{-1}\msf{N}_\Sigma(k)\right)^n(1+\msf{G}_0^R(k)\msf{D}_\Sigma(k))^{-1}.
\end{split}
\end{equation}
Since each component of $\msf{N}_\Sigma$ is of order $1/N$ (recall that $\Sigma(t_1,t_2)$ in Eq.\,\eqref{eq:self-energy} 
is proportional to $1/N$), it is sufficient to consider the first line in the RHS of Eq.\,\eqref{eq:GS dyson} 
for determining the Lyapunov exponent up to the order of $1/N$, 
\begin{equation}\label{eq:full RGF}
    G^R(k,\omega)\approx\frac{1}{\omega-\xi_k+\Sigma_0(k,\omega)}\approx\frac{1}{\omega-\xi_k+\Sigma_k},
\end{equation}  
where the self-energy correction $\Sigma_k$ is defined in Eq.\,\eqref{eq:self-energy frequency} or Eq.\,\eqref{eq:self-energy imag approx} in Appendix\,\ref{app:Lyapunov QMBS}.

\begin{figure}
    \centering
    \includegraphics*[width=0.7\linewidth]{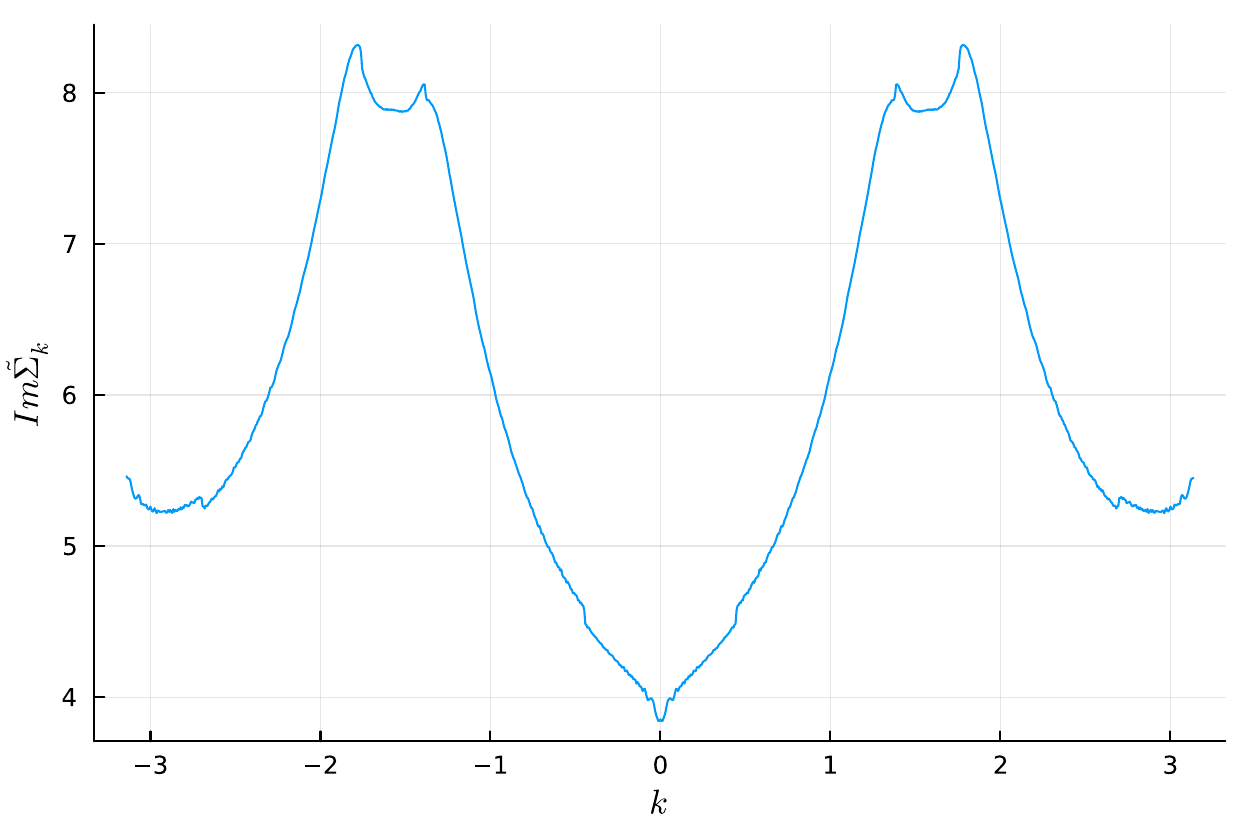}
    \caption{The imaginary part of the rescaled self-energy $\tilde{\Sigma}_k$ for each wavevector $k\in[-\pi,\pi]$. 
    For the parameters I set $\xi_k=-2\cos k+2.5$ and $L=2^{10}$. 
    The infinitesimal number $0^+$ in the definition of the retarded Green's function is set $0^+=3/2^{10}$.}
    \label{fig:self-energy imaginary}
\end{figure}

Note that the imaginary part of $\Sigma_k$ is more relevant than the real part for the 
LE since the real part merely yields a frequency shift. 
The expression of $\Im\Sigma_k$ in Eq.\,\eqref{eq:self-energy imag approx} shows that 
three scattering processes are relevant to the self-energy correction, and they represent either decay or amplification of the quasiparticles (Fig.\,\ref{fig:3 processes}):
the two-body scattering (the left diagram in Fig.\,\ref{fig:3 processes}) and the number increasing scattering (the middle diagram in Fig.\,\ref{fig:3 processes}) induce decay of the quasiparticles, 
while the number decreasing scattering (the right diagram in Fig.\,\ref{fig:3 processes}) might cause a dynamical instability. However, the numerical 
integration of $\Im\Sigma_k$ shown in Fig.\,\ref{fig:self-energy imaginary} indicates that the first two processes are always dominant. In Fig.\,\ref{fig:self-energy imaginary}, 
I plot the rescaled self-energy $\tilde{\Sigma}_k$ defined as, 
\begin{equation}
\Sigma_k=\frac{8J^2\alpha^4}{N}\tilde{\Sigma}_k.
\end{equation}
The positivity of $\Im\Sigma_k$ indicates that the quasi-particles always decay, meaning that at the single-particle level no chaos can be observed. This should be contrasted to 
standard QS's, 
where the bare Green's function already shows exponential growth.
\subsection{Bethe-Salpeter equation}
The absence of leading-order exponential growth in the Green's function implies that in order to capture the Lyapunov growth of the OTOC, 
the perturbative calculation requires the two-folded Keldysh contour, which was not necessary for the QS in Sec.\,\ref{sec:qs}. One can 
work out this perturbative expansion to find the following structure: the OTOC is written as, 
\begin{equation}\begin{split}
    C_\alpha(t)&=\frac{1}{NL}\sum_{k\in\mr{BZ}}f(k;(t,t),(0,0))\\
    C_\alpha(\omega)&=\frac{1}{NL}\sum_{k\in\mr{BZ}}\int\frac{dk_0}{2\pi}f(k,k_0,\omega),
\end{split}
\end{equation}    
where the second line is the Laplace transform of $C_\alpha(t)$. Indeed, at the zeroth order in $J$ (see Eq.\,\eqref{eq:OTOC zero qmbs}, also see Eq.\,\eqref{eq:zeroth order OTOC}) the OTOC satisfies this form by 
setting $f^{(0)}(k;(t,t),(0,0))=G^R_0(t,0)(G^R_0(t,0))^*$ and $f(k,\omega)=G^R_0(k,k_0)(G^R_0(k,k_0-\omega))^*$. 

\begin{figure}
    \centering
    \includegraphics*[width=0.7\linewidth]{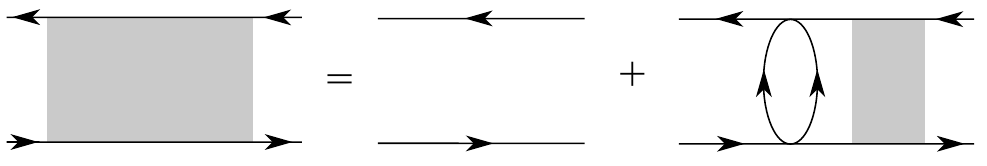}
    \caption{The Feynman diagram of the BS equation for the OTOC. The full correlator is obtained by resumming the ``rung'' diagram.}
    \label{fig:BS equation}
\end{figure}

The equation for $f(k;(t,t),(0,0))$ (Eq.\,\eqref{eq:2nd ord perturb OTOC} in Appendix\,\ref{app:Lyapunov QMBS}) is equivalent to the Bethe-Salpeter (BS) equation,
{
\begin{equation}\begin{split}
    f&(k,(t,t),(0,0))=f^{(0)}(k;(t,t),(0,0))\\&+\int_0^tdt_1dt_2f^{(0)}(k;(t,t),(t_1,t_2))\sum_{q\in\mr{BZ}}R(k,(t_1,t_2);q,(0,0))f(q;(t_1,t_2),(0,0)),
\end{split}
\end{equation}}
where $R$ is the ``rung diagram'' as shown in Fig.\,\ref{fig:BS equation}. More specifically, $R$ is written as, 
\begin{equation}
    R(k,(t_1,t_2);q,(0,0))=\frac{8J^2\alpha^4}{NL^2}\cos^2(Et_1)\cos^2(Et_2)K[G^W_0],
\end{equation}  
where $K[G^W_0]$ is a functional of the (bare) Wightman function $G^W_0$ (see Appendix.\,\ref{app:Lyapunov QMBS}).
One can further assume the on-shell condition $f(k,k_0,\omega)=\delta(k_0-\xi_k)f(k,\omega)$, which is an analog of the 
quasiparticle approximation in the Boltzmann equation, and obtain the following linear equation,
\begin{equation}\label{eq:BS equation main text}
    -i\omega f(k,\omega)\approx-\Im\Sigma_kf(k,\omega)+
    \sum_{n\in\mb{Z}}\sum_{q\in\mr{BZ}}K_{-n}(k,q)f(q,\omega+2nE),
\end{equation}
where the ``kernel'' $K_{-n}$ is defined in Appendix\,\ref{app:Lyapunov QMBS}. The LE corresponds to the 
eigenvalue of the RHS of Eq.\,\eqref{eq:BS equation main text}. Although solving this eigenvalue problem seems a formidable task, 
one can perform a similar perturbative calculation for the Green's function, exact up to order $1/N$. I give the detailed discussion 
in Appendix\,\ref{app:Lyapunov QMBS}. The upshot is that one needs to only consider $K_0$, i.e., 
\begin{equation}\label{eq:BS easy main text}
    -i\omega f(k,\omega)\approx-\Im\Sigma_kf(k,\omega)+\sum_{q\in\mr{BZ}}K_0(k,q)f(q,\omega).
\end{equation}
One can numerically compute the maximum eigenvalue of the RHS of Eq.\,\eqref{eq:BS easy main text}. For $L=2^{10}, \xi_k=-2\cos k+2.5$ 
I obtain
\begin{equation}
    \lambda\approx0.06567\frac{J^2\alpha^4}{N},
\end{equation} 
which is seen to vanish in the semiclassical limit, $N\to\infty$.

\section{Unstable periodic orbit from non-scar states}
\label{sec:unstable orbit}
\renewcommand{\theequation}{6.\arabic{equation}}
\setcounter{equation}{0}

The classical limit of the model [Eq.\,\eqref{eq:H}] hosts further periodic orbits: 
if one neglects the correlations ($\expval{a^2}\approx\expval{a}^2$ etc.), the following function satisfies the Heisenberg equation, 
\begin{equation}\label{eq:another orbit}
    \begin{split}
        \expval{a_{i\mu}(t)}=0,\,\,\expval*{b_{i\mu}(t)}=\beta_\mu e^{-iEt}.
    \end{split}
\end{equation}
Unlike the scar state $\ket*{\alpha}$ in Sec.\,\ref{sec:QMBS} where the corresponding solution satisfies the Heisenberg 
equation exactly, this solution describes approximate oscillations of the phase of $b$-bosons. Therefore, one expects that Eq.\,\eqref{eq:another orbit} should 
deviate from the exact dynamics after some time, which could be captured by studying an appropriate OTOC. 
In the following, I will show that this periodic solution 
is indeed unstable so that the Lyapunov exponent is of order unity.

\begin{figure}
    \centering
    \includegraphics[width=.65\linewidth]{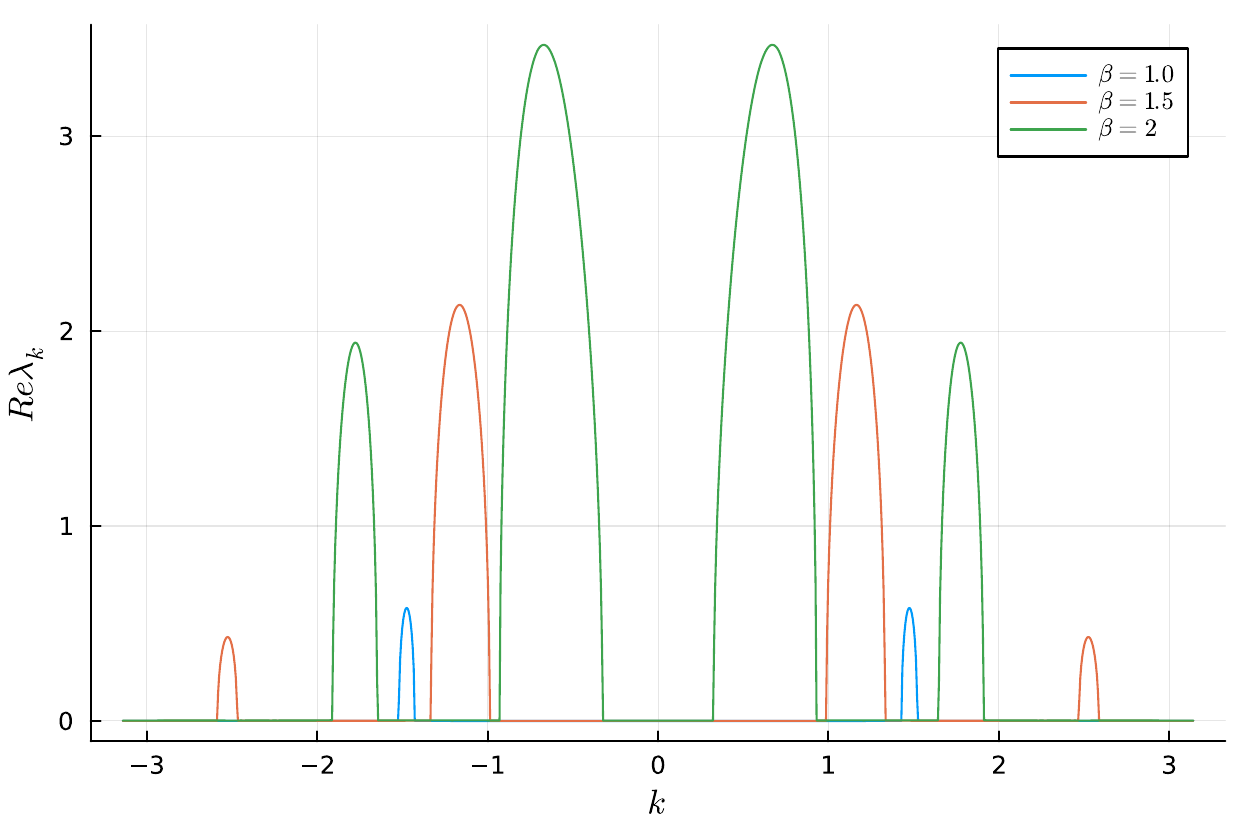}
    \caption{The largest real part of the Lyapunov exponent for each $k$ in the Brillouin zone for different $\beta$. 
    $\max_k\Re\lambda_k>0$ indicates an instability of the periodic orbit. I set $\xi_k=-2\cos k+2.5$, 
    $J=0.1$, and $L=2^{10}$.}
    \label{fig:lyapunov another orbit}
\end{figure}
For this periodic orbit I study the following OTOC $C_\beta(t)$,
\begin{equation}\label{eq:OTOC another orbit}
    C_\beta(t)=\frac{\theta(t)}{N^2}\sum_{i=1}^L\sum_{\mu,\nu=1}^N\expval*{[a_{i\mu}(t),a^\dag_{j\nu}(0)][a_{i\mu}(t),a^\dag_{j\nu}(0)]^\dag}{\beta},
\end{equation}
where the state $\ket*{\beta}$ corresponds to the classical solution Eq.\,\eqref{eq:another orbit},
\begin{equation}
    b_{i\mu}\ket*{\beta}=\beta_\mu\ket*{\beta},\,\,a_{i\mu}\ket*{\beta}=0.
\end{equation} 
To compute the OTOC I define the retarded Green's function $G^R$,
\begin{equation}\label{eq:another orbit GR}
    iG^R(k;t,t')=\theta(t-t')\expval*{[a_{k\mu}(t),a^\dag_{k\mu}(t')]}{\beta},
\end{equation}
which appears in the expression of the OTOC to leading order in the perturbation expansion, 
\begin{equation}
    C_\beta(t)=\frac{\theta(t)}{NL}\sum_{k\in\mr{BZ}}G^R(k;t,0)(G^R(k;t,0))^*.
\end{equation}
One can follow the same steps as in Sec.\,\ref{sec:qs} and find that the OTOC 
to the leading order can be written as,
\begin{equation}\label{eq:OTOC another orbit}
    C_\beta(t)=\frac{\theta(t)}{NL}\sum_{k\in\mr{BZ}}\left|\sum_\alpha e^{\lambda_{k\alpha}t}\left[u_{k\alpha}(t)u^\dag_{k\alpha}(0)\right]_{11}\right|^2,
\end{equation}
where $\Re\lambda_{k\alpha}(>0)$ is the LE of the unstable orbit (see Appendix\,\ref{app:OTOC another orbit} for the derivation). 

Fig.\,\ref{fig:lyapunov another orbit} shows the largest LE $\lambda_k\equiv\max_{\alpha=1,2}\Re\lambda_{k\alpha}$ for $k\in[-\pi,\pi]$ (note the ``band index'' $\alpha$ 
introduced in Appendix\,\ref{app:OTOC another orbit}). 
One can recognize several peaks of the LE centered around certain $k$'s in the Brillouin zone. Those depend on the initial condition 
$\ket*{\beta}$. This structure may be due to resonances between the periodic orbit and the dispersion $\xi_k$. Since $\max_k\Re\lambda_k>0$ in all cases, 
Fig.\,\ref{fig:lyapunov another orbit} indicates that the periodic orbits are unstable. 

It is important to note that the instability of this periodic orbit comes from the fact that the interaction 
$\sum_iv_i$ modifies the quadratic part of the Keldysh action such that the renormalized bare Green's function shows a dynamical instability. This should be contrasted 
to the previously discussed QMBS: due to the property $b_{i\mu}(t)\ket*{\psi}=0$, the interaction does not modify 
the quadratic action, which makes the LE of the scar state vanish in the semiclassical limit.


\section{Effect of a scar-breaking perturbation}
\label{sec:scar breaking perturbation}
\renewcommand{\theequation}{7.\arabic{equation}}
\setcounter{equation}{0}

Let us introduce the following perturbation to the Hamiltonian, 
\begin{equation}\label{eq:pert}
    V=-\frac{\epsilon}{2}\sum_{i=1}^L\sum_{\mu=1}^N\left(b^2_{i\mu}+(b^\dag_{i\mu})^2\right),
\end{equation}
which disrupts the SM structure. Consequently, the subspace $\mc{F}_a$ [Eq.\,\eqref{eq:scar subspace}] 
is no longer invariant under the perturbed Hamiltonian, leading to the eventual thermalization of the initial 
coherent state $\ket*{\alpha}$ [Eq.\,\eqref{eq:scar initial state}]. However, in the mean-field approximation, 
where the correlations are ignored, the periodic orbit [Eq.\,\eqref{eq:scar initial state} and 
Eq.\,\eqref{eq:scar evolution}] continues to satisfy the equation of motion. 
In the following, I study the stability of this orbit by computing the same OTOC $C_\alpha(t)$ [Eq.\,\eqref{eq:otoc QMBS}].

When the perturbation is sufficiently large, $|\epsilon|>E$, the quadratic component of the Hamiltonian has purely imaginary eigenvalues, indicating a dynamical instability. In this regime, the OTOC can be approximated as: 
\begin{equation}
    C_\alpha(t)\sim\frac{1}{NL}\sum_{\substack{k\in\mr{BZ}\\ \xi_k^2-\epsilon^2<0}}C_ke^{\sqrt{\epsilon^2-\xi_k^2}t},
\end{equation} 
where $C_k$ is a constant (see Appendix\,\ref{appsub:e>E} for the derivation). This mechanism mirrors that of the quasi-periodic motion in QS, where a perturbation alters the quadratic action, 
rendering certain modes unstable. 

\begin{figure}
    \centering
    \includegraphics[width=0.7\textwidth]{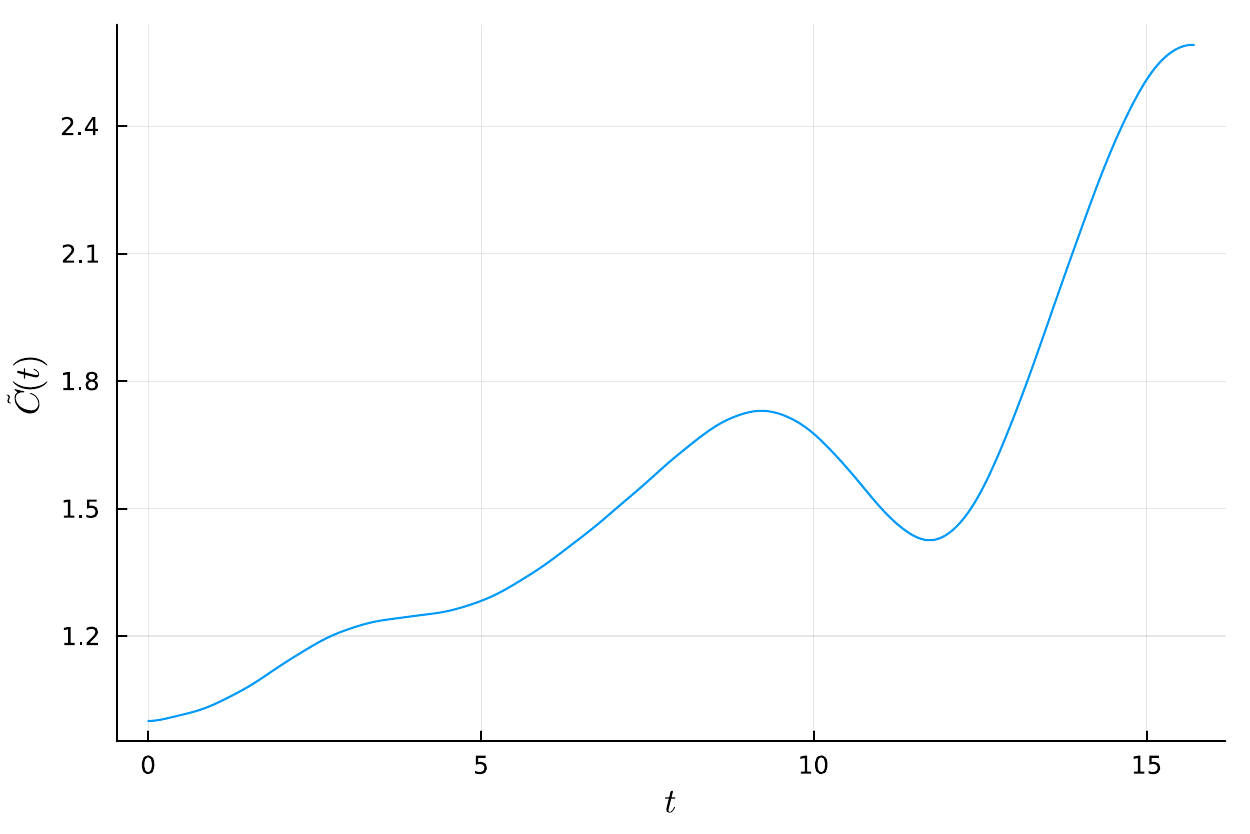}
    \caption{The rescaled OTOC $\tilde{C}(t)=NC_\alpha(t)$. I set $\xi_k=-2\cos k+2.5, 8J\alpha^2=3.0$, and $\epsilon=0.3$.}
    \label{fig:OTOC scar-breaking}
\end{figure}

For $|\epsilon|<E$, the quadratic action remains non-chaotic. Here, the first-order perturbative effects (Fig.\,\ref{fig:1st order}) become relevant, unlike for the exact QMBS. 
Solving the Dyson equation numerically (see Appendix.\,\ref{appsec:e<E} for the derivation), I find, however, that $C_\alpha(t)$ exhibits non-exponential growth for an early time period (Fig.\,\ref{fig:OTOC scar-breaking}). 
The absence of conventional Lyapunov growth stems from the fact that the interaction does not modify the quadratic part of the action, emphasizing the role of the non-linearity induced by the higher-order terms. 
Chaotic dynamics only emerge when higher-order correlations are included. 

A couple of brief remarks are in order. First, non-exponential growth of the OTOC under weak perturbations 
appears to be universal, regardless of the specific form chosen for $V$. I explored several other simple perturbations that preserve 
Eq.\,\eqref{eq:scar evolution} as an approximate solution of the equation of motion, and confirm that none of them leads to exponential 
OTOC growth. Second, similar subexponential growth has recently been observed in a classical PXP model, where deviations grow only polynomially 
over time when the initial state is two-site periodic\,\cite{MM_2024}. Since such weakly chaotic dynamics in the classical PXP model may be 
linked to the anomalous oscillations seen in the Rydberg experiment\,\cite{qmbs_experiment}, the mechanism discussed here 
could offer valuable insight into scarring phenomena beyond solvable toy models.


\section{Conclusion}
\label{sec:conclusion}
In this paper I discuss a connection between QMBS and classical chaos. The presented model serves as a counterexample to a 
widely-held conjecture that a proper semiclassical limit of QMBS should involve unstable periodic orbits. 
Through the lens of OTOCs, I instead demonstrate parametrically long 
coherent time when the number of flavors $N$ is sufficiently large.  
While the OTOC for the periodic orbit of the single-particle billiard system shows the conventional 
Lyapunov growth, the LE associated with the periodic orbit emanating from the scar states 
vanishes. Interestingly, this absence of exponential growth persists even under certain weak perturbations. 
In contrast, periodic orbits that are not connected to the scar states conform with the expected Lyapunov behavior, 
demonstrating the conventional chaotic dynamics in the classical limit.

The uncovered clear distinction between QMBS and classical chaos aligns with recent studies on a scarring phenomenon in many-body 
systems~\cite{PhysRevLett.130.250402,PhysRevLett.132.020401,PhysRevB.110.144302,pizzi2024} and an analysis on the PXP model in semiclassical regime~\cite{MM_2024}. 
The former demonstrated that certain eigenstates have enhanced overlap with unstable periodic orbits, while still satisfying the ETH. Therefore, 
unlike scar states in QMBS, the accurate analogs of the QS in many-body systems are in general thermal. Notably, Srednicki had already pointed out that the ETH 
is compatible with the QS in his seminal paper~\cite{eth_srednicki}, and these studies further confirm his insight in many-body contexts. 
The latter study discovered that a family of classical periodic orbits, which includes the experimentally relevant N\'eel orbit, is stable, exhibiting pure phase Lyapunov 
exponents (i.e., $|\lambda|=1$). This implies that non-chaotic long-lived oscillations in Rydberg atom chains originate from linearly stable 
periodic orbits, which contrasts sharply with the common speculation~\cite{qmbs_tvdp_prl} that such behaviors should be associated with instability of 
the periodic orbits.

The suppression of chaotic dynamics of OTOCs in QMBS stems from the SM construction. The interaction annihilates the scar states 
\textit{locally}, implying the absence of the Hartree term in perturbation theory. As the Hartree is
the only possible $O(1)$ contribution in the semiclassical limit $N\to\infty$, its absence precludes the emergence of chaos in these 
states. This is a distinctive feature of QMBS, not shared by general periodic orbits in the classical limit, 
as discussed in Sec.\,\ref{sec:qs} and Sec.\,\ref{sec:unstable orbit} where conventional Lyapunov growth is observed. 
Remarkably, even when a weak scar-breaking perturbation is introduced,
the anomalous OTOC dynamics persist, owning to the fact that the quadratic part of the action is not modified by the interaction. 
However, unlike the exact scar states, the higher-order
terms eventually induce chaos, driving the system toward thermalization.

These findings have practical implications for simulating or manipulating QMBS, either in laboratories or in quantum 
computers~\cite{PhysRevResearch.4.043027,Gustafson_2023}. The present model offers a solvable example of a 
QMBS that remains robust against weak scar-breaking perturbations, at least over a short timescale. Given that QMBS 
avoid the conventional chaotic dynamics even under a slight perturbation, it is plausible that signatures of QMBS can be detected in realistic systems, 
which may not perfectly preserve the fine-tuned structures of QMBS due to noise or environmental
decoherence. The findings of this paper suggest that the suppression of chaos may be a generic feature of QMBS in the semiclassical limit, including 
the PXP model. Ref.\,\cite{MM_2024} conducts the stability analysis of a classical PXP model, and finds that conventional Lyapunov 
growth is absent as well. Confirming the universality of this property in other models is an important step toward understanding the many-body 
scarring phenomenon. 


This study focuses on a weak coupling theory where the quasiparticles are well-defined. It is therefore interesting and worthwhile 
to explore similar analysis on strongly correlated large-$N$ theories that lack well-defined quasiparticles. For example, studying OTOCs of the spin-$S$ PXP model 
in semiclassical regime ($1\ll S<\infty$) may reveal important insight into its stability. Additionally, investigating the stability 
and chaos of more conventional and realistic models, such as the PXP model, is essential. While it might be challenging to study OTOCs 
and identify distinctive features in such models---due to the low dimensionality of their local Hilbert space, which often inhibits exponential growth of OTOCs  
even in chaotic systems~\cite{Kukuljan_2017}--- considering scar states as vacua of the SM projector $P_i$ (as defined in Eq.\,\eqref{eq:SM form}) 
offers a promising approach.

\section*{Acknowledgments}
I thank M. M\"uller for useful comments and thorough feedback on the manuscript. 
I am also indebted to C. Aron for stimulating discussions and teaching me 
the Keldysh path integral formalism through a doctoral course at EPFL.  
This work was supported by Grant No.\,200020200558 of the SNSF.


\appendix
\section{Derivation of the OTOC of the periodic orbit in the single-particle system}\label{app:OTOC QS}
\renewcommand{\theequation}{A\arabic{equation}}
\setcounter{equation}{0}
Here I provide a detailed derivation of the OTOC of QS [Eq.\,\eqref{eq:OTOC lyapunov qs}] in
the main text based on the Keldysh path integral approach. To construct the partition function I consider the following time evolution operator $\tilde{U}$,
\begin{equation}\label{eq:time evolution}
    \tilde{U}=U(\infty,0)U_0(0,-\infty), 
\end{equation}
where $U$ and $U_0$ satisfy the following differential equation
\begin{equation}
    i\hbar\pdv{t}U(t,0)=HU(t,0),\,\,i\hbar\pdv{t}U_0(t,0)=0,
\end{equation}
with the initial condition $U(0,0)=U_0(0,0)=1$. Physically, $\tilde{U}$ describes a sudden quench: until $t=0$ the system does not evolve at all, and then from $t=0$ 
the Hamiltonian $H$ acts on the system. While adding $U_0$ seems a redundant procedure due to $U_0(t,0)=U_0(0,t)=1$, $U_0$ removes the boundary term in the path integral and for this reason one should keep it. The partition function is defined as 
\begin{equation}\label{eq:action qs}\begin{split}
    Z&=\expval*{\tilde{U}^\dag\tilde{U}}{g}=\int D[a^*,a]e^{\frac{i}{\hbar}S_K}\\
    S_K&=\int_{-\infty}^\infty dt\sum_{\gamma=\pm}\gamma L_\gamma\\
    L_\gamma&=a^*_\gamma(t)i\dv{t}a_\gamma(t)-\theta(t)H[a_\gamma^*(t),a_\gamma(t)].
    \end{split}
\end{equation}
In order to consider the initial Gaussian wavepacket, the expectation value of the operator should be shifted, i.e., $\expval{a_\gamma}=\alpha_0(t=0)$. 
This requires that the complex field in the action should also be shifted as $a_\gamma\mapsto \alpha_0+a_\gamma$. Therefore the Lagrangian is re-written as 
\begin{equation}
    L_\gamma=a_\gamma^*(t)i\dv{t}a_\gamma(t)-\theta(t)H[\alpha^*_0(t)+a_\gamma^*(t),\alpha_0(t)+a_\gamma(t)].
\end{equation}
It is convenient to further rewrite the action in terms of the Keldysh basis $a_c$ and $a_q$, which are defined as 
\begin{equation}
    a_c(t)=\frac{1}{\sqrt{2}}(a_+(t)+a_-(t)),\,\,a_q(t)=\frac{1}{\sqrt{2}}(a_+(t)-a_-(t)).
\end{equation}
The (quadratic) leading part of the Lagrangian $L_{\mr{quad}}$ becomes
\begin{equation}\label{eq:quad action qs}\begin{split}
    &L_{\mr{quad}}=a^*_c(t)\left(i\dv{t}-i0^+\right)a_q(t)+a^*_q(t)\left(i\dv{t}+i0^+\right)a_c(t)\\
    &-\frac{\theta(t)}{2}\begin{pmatrix}
        a^*_c(t)&a_c(t)
    \end{pmatrix}M(t)\begin{pmatrix}
        a_q(t)\\ a^*_q(t)
    \end{pmatrix}
    -\frac{\theta(t)}{2}\begin{pmatrix}
        a^*_q(t)&a_q(t)
    \end{pmatrix}M(t)\begin{pmatrix}
        a_c(t)\\ a^*_c(t)
    \end{pmatrix},
    \end{split}
\end{equation}
where $M(t)$ is the monodromy matrix defined in Eq.\,\eqref{eq:monodromy}. 
Note that perturbations arise both from (the higher-order derivatives of) 
$H_0$ and (the whole) $H_n\,(n\geq1)$. However, $L_{\mr{quad}}$ is sufficient to calculate the OTOC to the leading order in $\hbar$, as shown below.

The Green's functions $\mc{G}^R, \mc{G}^A$, and $\mc{G}^K$ are defined as
\begin{equation}\begin{split}
    i\mc{G}^R(t,t')&=\expval{\begin{pmatrix}
        a_c(t)\\ a_c^*(t)
    \end{pmatrix}\begin{pmatrix}
        a_q^*(t')&a_q(t')
    \end{pmatrix}},\\
    i\mc{G}^A(t,t')&=\expval{\begin{pmatrix}
        a_q(t)\\ a_q^*(t)
    \end{pmatrix}\begin{pmatrix}
        a_c^*(t')&a_c(t')
    \end{pmatrix}},\\
    i\mc{G}^K(t,t')&=\expval{\begin{pmatrix}
        a_c(t)\\ a_c^*(t)
    \end{pmatrix}\begin{pmatrix}
        a_c^*(t')&a_c(t')
    \end{pmatrix}}.
    \end{split}
\end{equation}
From Eq.\,\eqref{eq:quad action qs} the bare retarded Green's function $\mc{G}^R_0$ is 
\begin{equation}\label{eq:bare G^R qs}
    \mc{G}_0^R(t,t')=\hbar\left(\left(i\dv{t}+i0^+\right)\sigma^3-\theta(t)M(t)\right)^{-1}(t,t'),
\end{equation}
where $\sigma^3=\mr{diag}(1,-1)$ is the Pauli matrix. The bare retarded Green's function $G_0^R$ in the main text [Eq.\,\eqref{eq:retarded QS}] is the $(1,1)$ component of $\mc{G}^R$. Eq.\,\eqref{eq:bare G^R qs} implies that the leading order of the OTOC [Eq.\,\eqref{eq:bare OTOC qs}] is $\hbar^2$.
Note that the same propagator can be obtained for $t,t'>0$ even if one removes the step function, i.e.,     
\begin{equation}\label{eq:approx G^R qs}
    \begin{split}
    \mc{G}_0^R(t,t')=\hbar\left(\left(i\dv{t}+i0^+\right)\sigma^3-M(t)\right)^{-1}(t,t'),
    \end{split}
\end{equation}
for $t,t'>0$. Indeed, Eq.\,\eqref{eq:bare G^R qs} and Eq.\,\eqref{eq:approx G^R qs} satisfy the same differential equation for $t,t'>0$.
Since $M(t)$ is periodic with period $T$ ($M(t+T)=M(t)$), calculating Eq.\,\eqref{eq:approx G^R qs} amounts to a non-interacting 
Floquet problem, which one can be solved formally as
\begin{equation}\label{eq:floquet G^R}
    \mc{G}^R_0(t,t')=-i\hbar\theta(t-t')\sum_\alpha\psi_\alpha(t)\psi^\dag_\alpha(t').
\end{equation}
$\psi_\alpha(t)$ is a Floquet normal mode with two components, satisfying 
\begin{equation}
    \left[\sigma^3\left(i\pdv{t}+i0^+\right)-M(t)\right]\psi_\alpha(t)=0.
\end{equation}
Just like in Bloch's theorem, the normal mode is decomposed as a phase factor and a periodic function,
\begin{equation}
    \psi_\alpha(t)=e^{\lambda_at}u_\alpha(t),\,\,u_\alpha(t+T)=u_\alpha(t),\,\,\,\,(\alpha=1,2).
\end{equation}
This phase fuctor is a LE of the periodic orbit. 

Finally, substituting Eq.\,\eqref{eq:floquet G^R} into Eq.\,\eqref{eq:bare OTOC qs}, the OTOC to the leading order is
\begin{equation}\label{eq:OTOC qs leading order}\begin{split}
    C^{(0)}(t)&=\hbar^2\theta(t)\left|\sum_\alpha e^{\lambda_\alpha t}\left[u_\alpha(t)u^\dag_\alpha(0)\right]_{11}\right|^2
    \sim\hbar^2e^{2\lambda t},
    \end{split}
\end{equation}
where $\lambda$ is the maximum among the real parts of the Lyapunov exponents, i.e., $\lambda=\max_\alpha(\Re\lambda_\alpha)$.

Corrections to Eq\,\eqref{eq:OTOC qs leading order} can be systematically computed by standard perturbation theory. 
For the leading order in the OTOC, the only important perturbation is the fourth order derivatives of 
$H_0$, i.e., $\phi^4$-like interactions. Such interactions can produce a Hartree diagram, which means that one has to solve 
a non-linear equation of motion for $G^R$. Thus such interactions might affect the LE even to the leading order. However, this correction 
is not important for the early time exponential growth of the OTOC. Furthermore, except for the Hartree term, 
one finds that perturbations only result in higher order corrections in $\hbar$ by simple power-counting: corrections from $H_n\,(n\geq1)$ are trivially higher order since 
the corresponding part of the action $S_n=-\int dtH_n$ contains at least $\hbar^0$ in the 
prefactor while each propagator carries $\hbar$. The higher order derivatives of the monodromy matrix in $H_0$ also yield only higher order corrections. 
Indeed, at each perturbative expansion, the overall prefactor gives $1/\hbar$ while there are at least two propagators ($\hbar^2$) involved.

\section{Derivation of thermal properties}\label{app:thermal derivation}
\renewcommand{\theequation}{B\arabic{equation}}
\setcounter{equation}{0}
In this Appendix I study thermodynamic properties of the many-body model, and 
prove that the density of the $b$-boson [Eq.\,\eqref{eq:b boson density}] should 
be strictly positive in the large-$N$ limit.

The action of the model is
{
\begin{equation}
    \begin{split}
        &S=\int_0^\beta d\tau\left[\sum_{k\in\mr{BZ}}\sum_{\mu=1}^Na^*_{k\mu}\left(\partial_\tau+\xi_k\right)a_{k\mu}+b^*_{k\mu}\left(\partial_\tau+\xi_k\right)b_{k\mu}+J\sum_{i=1}^L\sum_{\mu=1}^Nv_{ai}v_{bi}b^*_{i\mu}b_{i\mu}\right]\\
        &+\int_0^\beta d\tau N\sum_{i=1}^L\left[\frac{\Delta_{ai}}{2}\left(v_{ai}-\frac{1}{N}\sum_{\mu=1}^N(a_{i\mu}+a^*_{i\mu})^2\right)+\frac{\Delta_{bi}}{2}\left(v_{bi}-\frac{1}{N}\sum_{\mu=1}^N(b_{i\mu}+b_{i\mu}^*)^2\right)\right],
    \end{split}
\end{equation}}
where $\Delta_{a/bi}$ is a Lagrange multiplier. Assuming that $\Delta_{a/b}$ and $v_{a/b}$ is space-time independent, 
the action becomes
{
\begin{equation}\label{eq:action}
    \begin{split}
        S|_{\Delta,v:\mr{fixed}}&=\frac{N}{2}\sum_{k\in\mr{BZ}}\sum_{i\omega_n}\mr{tr}\log
        \begin{pmatrix}
            -i\omega_n+\xi_k-\Delta_a & -\Delta_a\\ -\Delta_a&i\omega_n+\xi_{-k}-\Delta_a
        \end{pmatrix}
        \\
        &+\frac{N}{2}\sum_{k\in\mr{BZ}}\sum_{i\omega_n}\mr{tr}\log
        \begin{pmatrix}
            -i\omega_n+\xi_k+Jv_av_b-\Delta_b & -\Delta_b\\ -\Delta_b&i\omega_n+\xi_{-k}+Jv_av_b-\Delta_b
        \end{pmatrix}\\
        &+\frac{NL}{2T}\left(\Delta_av_a+\Delta_bv_b\right),
    \end{split}
\end{equation}}
where $\omega_n=2\pi nT$ is the Matsubara frequency.
The minimizing the action with respect to $\Delta_{a/b}$ and $v_{a/b}$ yields the mean-field equations,
\begin{equation}\label{eq:mfeq}
    \begin{split}
        \Delta_a&=-\frac{Jv_b}{L}\sum_{k\in\mr{BZ}}\frac{\xi_k+Jv_av_b-\Delta_b}{E_{kb}}\coth(\frac{E_{kb}}{2T})\\
        \Delta_b&=-\frac{Jv_a}{L}\sum_{k\in\mr{BZ}}\frac{\xi_k+Jv_av_b-\Delta_b}{E_{kb}}\coth(\frac{E_{kb}}{2T})\\
        v_a&=\frac{1}{L}\sum_{k\in\mr{BZ}}\frac{\xi_k}{E_{ka}}\coth(\frac{E_{ka}}{2T})\\
        v_b&=\frac{1}{L}\sum_{k\in\mr{BZ}}\frac{\xi_k+Jv_av_b}{E_{kb}}\coth(\frac{E_{kb}}{2T}),
    \end{split}
\end{equation}
where the quasiparticle energy for the two bosonic modes, $E_{ka}$ and $E_{kb}$, are defined as
\begin{equation}\begin{split}
    E_{ka}&=\sqrt{\xi_k(\xi_k-2\Delta_a)}\\
    E_{kb}&=\sqrt{(\xi_k+Jv_av_b)(\xi_k+Jv_av_b-2\Delta_b)}.
    \end{split}
\end{equation}

Multiplying the action Eq.\,\eqref{eq:action} by the temperature provides the free energy density. While the energy density can also 
be obtained by differentiating the action with respect to the temperature, one can instead explicitly evaluate the thermal 
average of the Hamiltonian with the mean-field action, which results in the following expression,
\begin{equation}\label{eq:free energy}\begin{split}
    \frac{F}{NL}&=\frac{1}{\beta L}\sum_{k\in\mr{BZ}}\left[\log\left(e^{E_{ka}/2T}-e^{-E_{ka}/2T}\right)+\log\left(e^{E_{kb}/2T}-e^{-E_{kb}/2T}\right)\right]\\&+\frac{1}{2}\left(\Delta_av_a+\Delta_bv_b\right)\\
    \frac{E}{NL}&=\frac{1}{L}\sum_{k\in\mr{BZ}}\frac{\xi_k(\xi_k-\Delta_a)}{2E_{ka}}\coth(\frac{E_{ka}}{2T})\\
    &+\frac{1}{L}\sum_{k\in\mr{BZ}}\left[\frac{(\xi_k+Jv_av_b)(\xi_k+Jv_av_b-\Delta_b)}{2E_{kb}}\coth(\frac{E_{kb}}{2T})-\xi_k\right]-\frac{Jv_av_b}{2},
    \end{split}
\end{equation}
where $T$ is the temperature.

\begin{figure}
    \begin{minipage}{0.48\linewidth}
        \includegraphics[width=\linewidth]{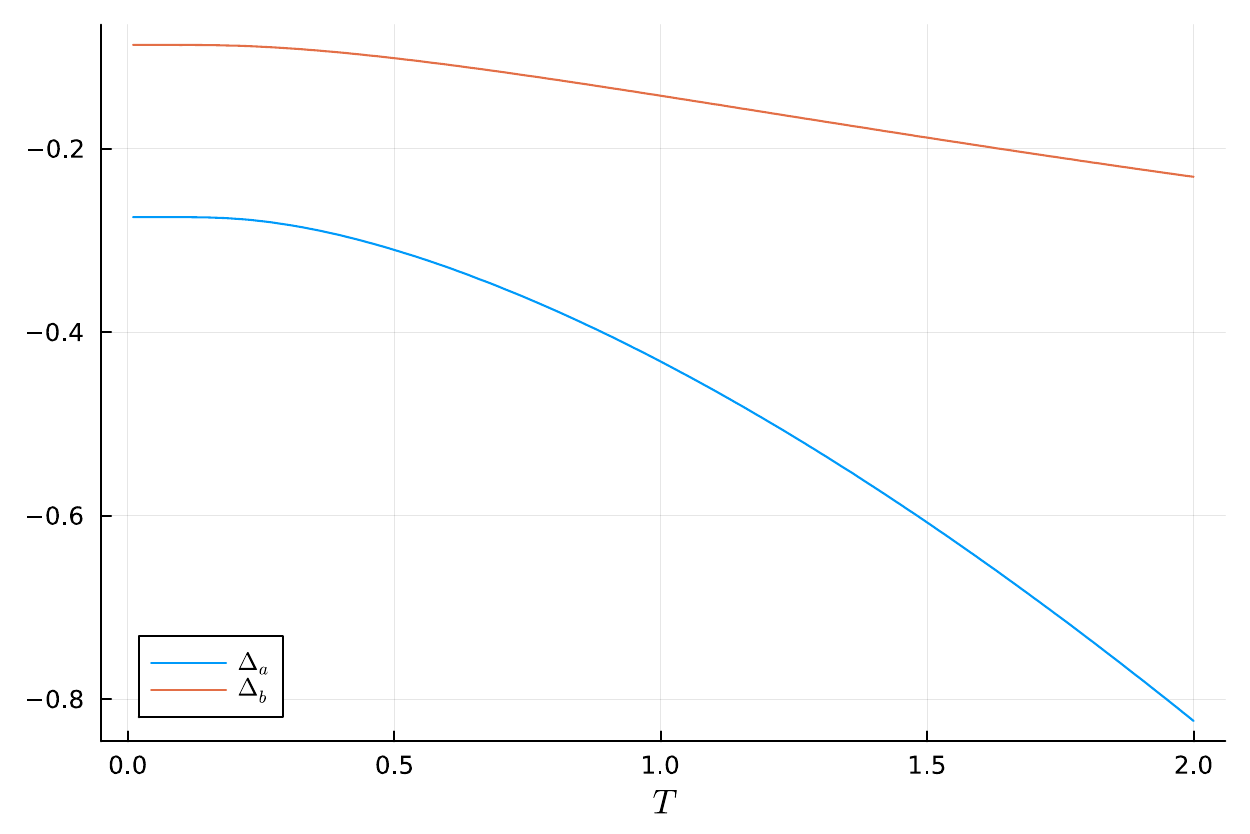}
    \end{minipage}
    \begin{minipage}{0.48\linewidth}
        \includegraphics[width=\linewidth]{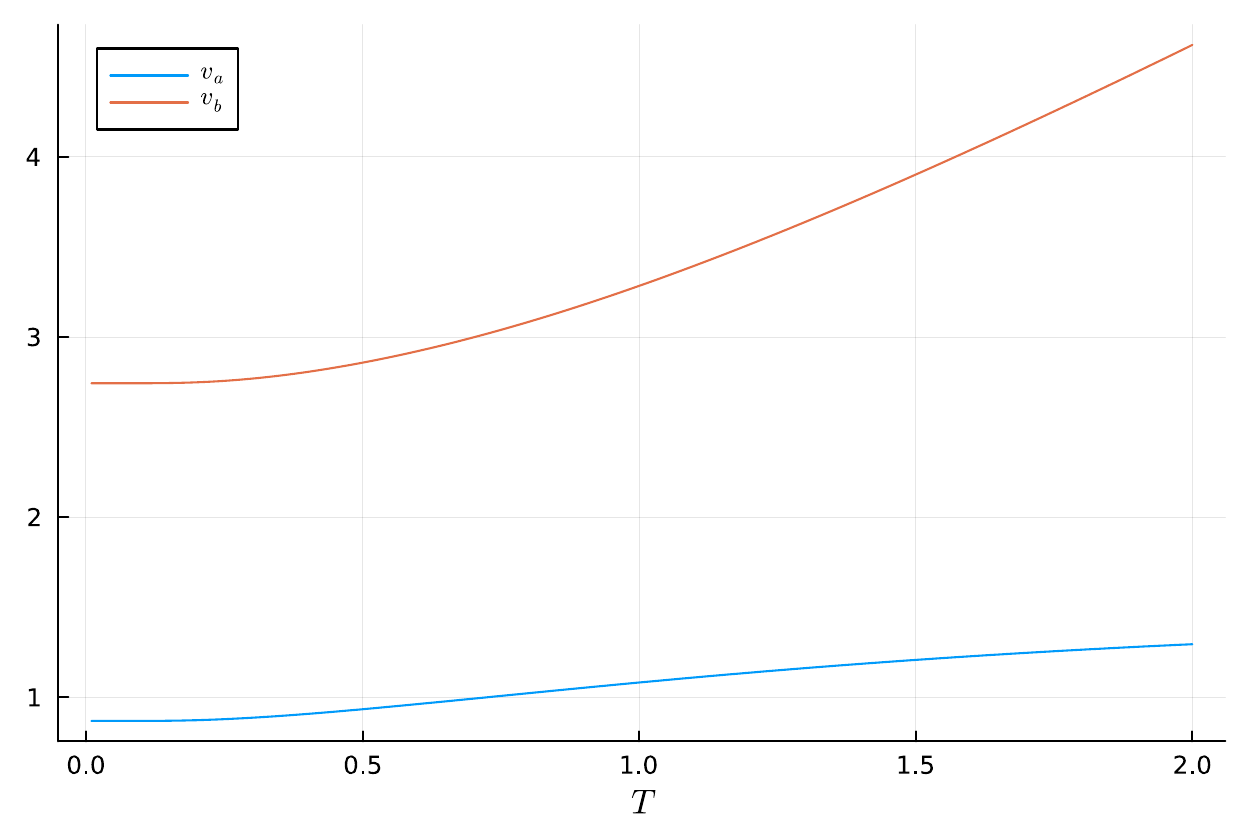}
    \end{minipage}
    \caption{Solution of the mean-field equation Eq.\,\eqref{eq:mfeq} as functions of the temperature $T$.}
\end{figure}

Finally I will show that Eq.\,\eqref{eq:mfeq} implies non-zero $b$-boson density [Eq.\,\eqref{eq:b boson density}] at any temperature 
by contradiction.
The mean-field equation implies that the thermal expectation value of $\rho_b$ satisfies the following relation,
\begin{equation}\label{eq:density relation}
    \expval*{\rho_b}=-\frac{\Delta_a}{2Jv_b}=-\frac{\Delta_b}{2Jv_a}.
\end{equation}
Suppose $\expval*{\rho_b}=0$. By Eq.\,\eqref{eq:density relation} it implies $\Delta_a=\Delta_b=0$. The first two lines of Eq.\,\eqref{eq:mfeq} implies
\begin{equation}
    0=-\frac{Jv_{a/b}}{L}\sum_{k\in\mr{BZ}}\mr{sgn}(\xi_k+Jv_av_b)\coth(\frac{E_{kb}}{2}).
\end{equation}
This in turns implies that $v_b$ should also be zero by the last line of Eq.\,\eqref{eq:mfeq} because it reads
\begin{equation}\label{eq:vb proof}
    v_b=\frac{1}{L}\sum_{k\in\mr{BZ}}\mr{sgn}(\xi_k+Jv_av_b)\coth(\frac{\beta E_{kb}}{2}),
\end{equation}
which indicates the relation $\Delta_a=-Jv_b^2=0$. However, if $v_b=0$, the right-hand side of Eq.\,\eqref{eq:vb proof} implies
\begin{equation}
    0=v_b=\frac{1}{L}\sum_{k\in\mr{BZ}}\mr{sgn}(\xi_k)\coth(\frac{\beta|\xi_k|}{2})>0,
\end{equation}
which is a contradiction.
\section{Derivation of the Lyapunov exponent in QMBS}
\label{app:Lyapunov QMBS}
\renewcommand{\theequation}{C\arabic{equation}}
\setcounter{equation}{0}
In this Appendix I provide a detailed derivation of the BS equation [Eq.\,\eqref{eq:BS equation main text}] 
in the main text based on the Keldysh path integral formalism. Following the main text I first derive the self-energy correction to the 
regarded Green's function in Sec.\,\ref*{app:retarded}, and then derive the BS equation in Sec.\,\ref{app:BS eq}.

The calculation below relies on the non-equilibrium Green's functions defined as,
\begin{equation}\label{eq:Green function}
    \begin{split}
        iG^R(k;t,t')&=\theta(t-t')\expval*{[b_{k\mu}(t),b^\dag_{k\mu}(t')]}{\alpha}\\
        iG^A(k;t,t')&=-\theta(t-t')\expval*{[b_{k\mu}(t),b^\dag_{k\mu}(t')]}{\alpha}\\
        iG^K(k;t,t')&=\expval*{\{b_{k\mu}(t),b^\dag_{k\mu}(t')\}}{\alpha}.
    \end{split}
\end{equation}
\subsection{The regarded Green's function}
\label{app:retarded}
The partition function of the system is written as 
\begin{equation}\begin{split}
    Z&=\expval*{U^\dag U}{\alpha}=\int D[a^*,a]D[b^*,b]e^{iS_K}\\
    S_K&=\int_{-\infty}^\infty dt\sum_{\gamma=\pm}\gamma L_\gamma\\
    L_\gamma&=\sum_{k\in\mr{BZ}}\sum_{\mu=1}^Na^*_{k\mu\gamma}(t)i\dv{t}a_{k\mu\gamma}(t)+b^*_{k\mu\gamma}(t)i\dv{t}b_{k\mu\gamma}(t)-H_\gamma,
\end{split}
\end{equation}
where $H_\gamma$ is the Hamiltonian with $a_\gamma$ and $b_\gamma$ fields. The equation of motion for those fields are
\begin{equation}\label{eq:eom Keldysh}
    \begin{split}
        i\dv{a_{k\mu\gamma}}{t}&=\frac{\delta H_\gamma}{\delta a^*_{k\mu\gamma}(t)},\,\,
        i\dv{b_{k\mu\gamma}}{t}=\frac{\delta H_\gamma}{\delta b^*_{k\mu\gamma}(t)}.
    \end{split}
\end{equation}
A solution of this equation is 
\begin{equation}\label{eq:scar solution}
    a_{k\mu\gamma}(t)=\delta_{k,0}\alpha_\mu e^{-iEt},\,\,b_{k\mu\gamma}(t)=0.
\end{equation}
This solution corresponds to the perpetual oscillations of the initial scar state $\ket*{\alpha}$.  
One can study the dynamics of the fluctuation by shifting $a_\gamma$ field as 
$a_\gamma\mapsto a_\gamma+\alpha_\gamma$. I define the interacting term $L_{\mr{int}}=\sum_{\gamma=\pm}\gamma L_{\mr{int},\gamma}$ as follows,
{
\begin{equation}\label{eq:L_int}
    \begin{split}
        L_{\mr{int},\gamma}&=-\frac{J}{N^2}\sum_{i=1}^L\sum_{\mu\nu\rho=1}^N\left(4\alpha_\mu^2\cos^2(Et)+4\alpha_\mu\cos(Et)(a_{i\mu\gamma}+a^*_{i\mu\gamma})
        +(a_{i\mu\gamma}+a_{i\mu\gamma}^*)^2\right)\\&\qquad\qquad\times(b_{i\nu\gamma}+b_{i\nu\gamma}^*)^2b^*_{i\rho\gamma}b_{i\rho\gamma}\\
        &\approx-\frac{4J\alpha^2}{N}\cos^2(Et)\sum_{i=1}^L\sum_{\mu\nu=1}^Nb_{i\mu\gamma}^*b_{i\mu\gamma}(b_{i\nu\gamma}+b_{i\nu\gamma}^*)^2,
    \end{split}
\end{equation}} 
where the approximation in the second line is valid under the assumption that $\alpha^2=N^{-1}\sum_\mu\alpha_\mu^2$ is sufficiently large.

It is now convenient to rewrite $L_{\mr{int}}$ in the Keldysh basis ($b_c$ and $b_q$),
{
    \begin{equation}
        \begin{split}
            L_{\mr{int}}&\approx-\frac{2J\alpha^2}{N}\cos^2(Et)\sum_{i=1}^L\sum_{\mu\nu=1}^N\left(b^*_{i\mu c}b_{i\mu q}+b^*_{i\mu q}b_{i\mu c}\right)
            \left((b_{i\nu c}^*)^2+(b^*_{i\nu q})^2+b_{i\nu c}^2+b_{i\nu q}^2\right)\\
            &-\frac{4J\alpha^2}{N}\cos^2(Et)\sum_{i=1}^L\sum_{\mu\nu=1}^N\left(b^*_{i\mu c}b^*_{i\mu q}+b_{i\mu c}b_{i\mu q}\right)
            \left(b^*_{i\nu c}b_{i\nu c}+b^*_{i\nu q}b_{i\nu q}\right)\\
            &-\frac{8J\alpha^2}{N}\cos^2(Et)\sum_{i=1}^L\sum_{\mu\nu=1}^N\left(b^*_{i\mu c}b_{i\mu q}+b^*_{i\mu q}b_{i\mu c}\right)
            \left(b^*_{i\nu c}b_{i\nu c}+b^*_{i\nu q}b_{i\nu q}\right).
        \end{split}
    \end{equation}}
    The retarded Green's function is then defined as $iG^R(t,t')=\expval*{b_c(t)b_q^*(t')}$.
    As stated in the main text, the first order perturbation (the Hartree term) should vanish because the scar
states are annihilated by $b_{i\mu}$. Up to the order of $1/N$, the second order perturbative correction to the retarded Green's function yields 
\begin{equation}
    -\int_{-\infty}^\infty dt_1dt_2G_0^R(k;t,t_1)\frac{8J^2\alpha^4}{N}\cos^2(Et_1)\cos^2(Et_2)\tilde{\Sigma}[G_0]G_0^R(k;t_2,0),
\end{equation}
where 
\begin{equation}
    \label{eq:Sigma functional}\begin{split}
    \tilde{\Sigma}[G_0]&=\frac{1}{L^2}\sum_{p,q\in\mr{BZ}}G_0^R(p;t_1,t_2)G_0^R(q;t_1,t_2)G^R_0(k-p-q;t_1,t_2)\\
    &+\frac{3}{L^2}\sum_{p,q\in\mr{BZ}}G_0^R(p;t_1,t_2)G^K_0(q;t_1,t_2)G^K_0(k-p-q;t_1,t_2)\\
    &+\frac{16}{L^2}\sum_{p,q\in\mr{BZ}}G^R_0(p;t_1,t_2)G_0^K(q;t_1,t_2)G_0^K(-k+p+q;t_2,t_1)\\
    &+\frac{2}{L^2}\sum_{p,q\in\mr{BZ}}G^R_0(p;t_1,t_2)G^K_0(q;t_2,t_1)G_0^K(p-k-q;t_2,t_1)\\
    &+\frac{8}{L^2}\sum_{p,q\in\mr{BZ}}G^A_0(p;t_2,t_1)G^K_0(q;t_1,t_2)G_0^K(k+p-q;t_1,t_2)\\
    &+\frac{6}{L^2}\sum_{p,q\in\mr{BZ}}G^A_0(p;t_2,t_1)G_0^K(q;t_2,t_1)G_0^K(k+p+q;t_1,t_2).
    \end{split}
\end{equation}
In the frequency space, the Dyson equation is written as, 
\begin{equation}
    G^R(k,\omega)=G^R_0(k,\omega)-G^R_0(k,\omega)\sum_{n\in\mb{Z}}\Sigma_n(k,\omega)G^R(k,\omega+2nE),
\end{equation}
where 
\begin{equation}\label{eq:self-energy frequency}
    \begin{split}
        \Sigma_0(k,\omega)&=-\frac{8J^2\alpha^4}{NL^2}\sum_{p,q\in\mr{BZ}}\left(S^R(\omega,k,p,q)+\frac{1}{4}S^R(\omega+2E,k,p,q)+\frac{1}{4}S^R(\omega-2E,k,p,q)\right)\\
        \Sigma_{\pm1}(k,\omega)&=-\frac{4J^2\alpha^4}{NL^2}\sum_{p,q\in\mr{BZ}}\left(S^R(\omega,k,p,q)+S^R(\omega\pm2E,k,p,q)\right)\\
        \Sigma_{\pm2}(k,\omega)&=-\frac{2J^2\alpha^4}{NL^2}\sum_{p,q\in\mr{BZ}}S^R(\omega\pm2E,k,p,q)\\
        \Sigma_{\pm n}(k,\omega)&=0\,\,(n>2).
    \end{split}
\end{equation}
The factor $S^R$ is defined as 
\begin{equation}\begin{split}
    S^R(\omega,k,p,q)&=\frac{1}{\omega-\xi_p-\xi_q-\xi_{k-p-q}+i0^+}+\frac{2}{\omega+\xi_p-\xi_q-\xi_{k+p-q}+i0^+}\\
    &-\frac{1}{\omega+\xi_p+\xi_q-\xi_{k+p+q}+i0^+}.
\end{split}
\end{equation}

The imaginary part of the self-energy $\Im\Sigma_n(k,\omega)$ is approximated by substituting   
\begin{equation}\label{eq:self-energy imag approx}
    \begin{split}
        S^R(\omega,k,p,q)&\rightarrow-\pi\left[\delta(\xi_k-\xi_p-\xi_q-\xi_{k-p-q})+2\delta(\xi_k+\xi_p-\xi_q-\xi_{k+p-q})\right.\\
        &\left.\qquad-\delta(\xi_k+\xi_p+\xi_q-\xi_{k+p+q}+i0^+)\right].
    \end{split}
\end{equation}
into Eq.\,\eqref{eq:self-energy frequency}. The first two terms in $S^R$ represent the particle-number decreasing process  
and the number conserving process, respectively, and the last term is the number increasing process. 
Combined with Eq.\,\eqref{eq:full RGF} the approximate retarded Green's function is 
\begin{equation}
    G^R(k,\omega)\approx\frac{1}{\omega-\xi_k+\Sigma_k},
\end{equation}
where
\begin{equation}\begin{split}\label{eq:self-energy simple}
    \Sigma_k&=-\frac{8J^2\alpha^4}{NL^2}\sum_{p,q\in\mr{BZ}}\left(S^R(\xi_k,k,p,q)+\frac{1}{4}S^R(\xi_k+2E,k,p,q)+\frac{1}{4}S^R(\xi_k-2E,k,p,q)\right)\\
    &\equiv\frac{8J^2\alpha^4}{N}\tilde{\Sigma}_k.
\end{split}
\end{equation}
\subsection{Bethe-Salpeter equation}
\label{app:BS eq}

In this section I will derive the BS equation in the main text\,[Eq.\,\eqref{eq:BS equation main text}], following equilibrium cases\,\cite{Stanford_2016,Patel_2017}.
Unlike the Green's function, one needs to double the degrees of freedom to regard the OTOC as a path-ordered correlator. 
The partition function with the two folded Keldysh contour is,
\begin{equation}
    \begin{split}
        Z&=\int D[a^*,a]D[b^*,b]e^{iS_K}\\
        S_K&=\sum_{p=u,d}\sum_{\gamma=\pm}\gamma S_{p\gamma}\\
        S_{p\gamma}&=\int_{-\infty}^\infty dt\sum_{k\in\mr{BZ}}\sum_{\mu=1}^N
        \left[a^*_{k\mu p\gamma}(t)i\dv{t}a_{k\mu p\gamma}(t)
        +b^*_{k\mu p\gamma}(t)i\dv{t}b_{k\mu p\gamma}(t)\right]
        -H_{p\gamma}.
    \end{split}
\end{equation} 
To carry out perturbative calculations I follow the idea of sudden quench in the single-particle case: 
since one only needs to consider the system for the period $0\leq\tau\leq t$, it is sufficient to introduce 
the interaction term in the action for this period, i.e., $S_{\mr{int}}$ is defined as 
\begin{equation}\label{eq:S_int}
    S_{\mr{int}}=\int_0^td\tau L_{\mr{int}}.
\end{equation} 
This way of introducing the interaction does not affect the final result, and simplifies calculations since 
one can take advantage of the causality (any field $\phi(\tau)$ in $S_{\mr{int}}$ should satisfy $0\leq\tau\leq t$).

In order to derive the BS equation for the OTOC, I consider the Laplace transform 
of $C_\alpha(t)$ as 
\begin{equation}\begin{split}
    C_\alpha(\omega)&=\int_0^\infty dt e^{i\omega t}C_\alpha(t)\\
    C_\alpha(t)&=\int\frac{d\omega}{2\pi}e^{-i\omega t}C_\alpha(\omega),
\end{split}
\end{equation} 
where for the inverse transformation the integral contour should run above all the 
singularities. Note that the conventional Fourier transformation is insufficient since 
the exponential function cannot be Fourier transformed. In the following I will compute $C_\alpha(\omega)$ perturbatively, and 
derive the BS equation for the OTOC.  

\subsubsection{The zeroth order}\label{appsec:zeroth order}
The zeroth order of the OTOC is already presented in the main text [Eq.\,\eqref{eq:OTOC zero qmbs}]. 
Its Laplace transform is 
\begin{equation}\label{eq:zeroth order OTOC}
    C^{(0)}_\alpha(\omega)=\frac{1}{NL}\int\frac{dk_0}{2\pi}\sum_{k\in\mr{BZ}}G_0^R(k,k_0)\left(G^R_0(k,k_0-\omega)\right)^*
    \equiv\frac{1}{NL}\int\frac{dk_0}{2\pi}\sum_{k\in\mr{BZ}}f^{(0)}(k,k_0,\omega),
\end{equation}
where $k_0$ runs above all the singularities too. When higher order corrections are considered in Sec.\,\ref{appsec:higher order}, Eq.\,\eqref{eq:zeroth order OTOC} 
should be replaced with the full propagators. However, such a term is irrelevant for the Lyapunov growth of the OTOC, 
since $\Sigma_k$ [Eq.\,\eqref{eq:self-energy simple}] is positive, which implies that Eq.\,\eqref{eq:zeroth order OTOC} has a pole 
in the lower half plane and thus $C_\alpha^{(0)}(t)$ decays in time.

\subsubsection{The first order}
As stated in the main text in Sec.\,\ref{subsec:1st order} (or Sec.\,\ref{app:retarded}), the first order perturbation should vanish, as 
any normal ordered operator should annihilate the scar states. 

\subsubsection{The second order}
\begin{figure}
    \begin{minipage}{0.48\linewidth}
    \includegraphics[width=\linewidth]{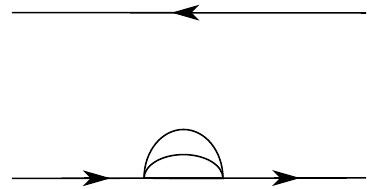}
    \end{minipage}
    \begin{minipage}{0.48\linewidth}
        \includegraphics[width=\linewidth]{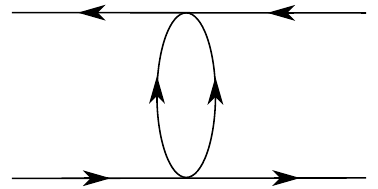}
    \end{minipage}
    \caption{Typical diagrams of the second order corrections to the OTOC. They include (i) the correction 
    to the propagator (the left panel) and (ii) the one peculiar to the OTOC (the right panel).}
    \label{fig:otoc digram}
\end{figure}
The second order perturbations are classified as (i) perturbations to the Green's function (the left panel of Fig.\,\ref{fig:otoc digram}) and (ii) a perturbation
peculiar to the OTOC (the right panel of Fig.\,\ref{fig:otoc digram}). The perturbations (i) have been already discussed in Sec.\ref{app:retarded}, so I focus on the 
perturbation (ii). Unlike the correction to the Green's function that contains many terms [Eq.\,\eqref{eq:Sigma functional}], 
the perturbation (ii) consists of only one term, 
{
\begin{equation}\label{eq:2nd ord perturb OTOC}
    \begin{split}
        &\int_0^t dt_1dt_2\frac{1}{NL}\sum_{k\in\mr{BZ}}G^R_0(k;t,t_1)(G^R_0(k;t,t_2))^*\frac{8J^2\alpha^4}{NL^2}\cos^2(Et_1)\cos^2(Et_2)\\
        &\qquad\times\sum_{p,q\in\mr{BZ}}G_0^W(p;t_1,t_2)G_0^W(k-p-q;t_1,t_2)
        G^R_0(q;t_1,0)(G^R_0(q;t_2,0))^*,
    \end{split}
\end{equation}} 
where $G^W_0$ is the zeroth order approximation of (an analog of) the Wightman function $G^W(t,t')=\expval*{b(t)b^\dag(t')}{\alpha}$.
As shown in Fig.\,\ref{fig:otoc digram}, the second order correction to the OTOC corresponds to adding a ``rung diagram'' to the set of 
the bare propagators. The domain of the integral can now be extended from $[0,t]$ to $[-\infty,\infty]$ due to the causality structure of the retarded 
Green's function ($G^R(t,t')\propto\theta(t-t')$).

In the frequency space, the above equation is written as,
\begin{equation}\label{eq:2nd order OTOC}\begin{split}
    \int&\frac{dk_0dq_0}{(2\pi)^2}\frac{1}{NL}\sum_{k\in\mr{BZ}}f^{(0)}(k,k_0,\omega)
    \sum_{n\in\mb{Z}}\sum_{q\in\mr{BZ}}R_n(k,k_0;q,q_0)f^{(0)}(q,q_0,\omega+2nE),
\end{split}
\end{equation}
where $R_n$ is defined as 
\begin{equation}
    \begin{split}
        R_0(k,k_0;q,q_0)&=\frac{2J^2\alpha^4}{NL^2}\sum_{p\in\mr{BZ}}G_0^W(k-p-q,k_0-q_0-\xi_p)\\
        &+\frac{J^2\alpha^4}{2NL^2}\sum_{p\in\mr{BZ}}G_0^W(k-p-q,k_0-q_0-\xi_p+2E)\\
        &+\frac{J^2\alpha^4}{2NL^2}\sum_{p\in\mr{BZ}}G_0^W(k-p-q,k_0-q_0-\xi_p-2E)\\
        R_{\pm1}(k,k_0;q,q_0)&=\frac{J^2\alpha^4}{NL^2}\sum_{p\in\mr{BZ}}G_0^W(k-p-q,k_0-q_0-\xi_p)\\
        &+\frac{J^2\alpha^4}{NL^2}\sum_{p\in\mr{BZ}}G^W_0(k-p-q,k_0-q_0-\xi_p\pm2E)\\
        R_{\pm2}(k,k_0;q,q_0)&=\frac{J^2\alpha^4}{2NL^2}\sum_{p\in\mr{BZ}}G_0^W(k-p-q,k_0-q_0-\xi_p\pm2E)\\
        R_{\pm n}(k,k_0;q,q_0)&=0\,\,(n>2).
    \end{split}
\end{equation}

\subsubsection{Higher order}\label{appsec:higher order}
The above exercise (Eq.\,\eqref{eq:zeroth order OTOC} and Eq.\,\eqref{eq:2nd order OTOC}) implies 
that the higher order perturbations correspond to adding many rung diagrams and thus they can be incorporated 
by replacing the bare correlator with the full correlator: let us write the OTOC in the 
momentum space as 
\begin{equation}
    C_\alpha(\omega)=\int\frac{dk_0}{2\pi}\frac{1}{NL}\sum_{k\in\mr{BZ}}f(k,k_0,\omega).
\end{equation}
At the zeroth order in $J$, $f(k,k_0,\omega)=f^{(0)}(k,k_0,\omega)+O(J)$. The effect of higher order corrections is modifying 
the Green's function and adding the rung diagram. 
Therefore, $f(k,k_0,\omega)$ satisfies the following BS equation,
\begin{equation}\label{eq:BS second to last}\begin{split}
    f(k,k_0,\omega)&=G^R(k,k_0)(G^R(k,k_0-\omega))^*\\
    +&G^R(k,k_0)(G^R(k,k_0-\omega))^*\int\frac{dq_0}{2\pi}\sum_{n\in\mb{Z}}\sum_{q\in\mr{BZ}}R(k,k_0;q,q_0)f(q,q_0,\omega+2nE).
\end{split}
\end{equation}
As mentioned in Sec.\,\ref{appsec:zeroth order}, one can safely neglect the first term in the RHS for the asymptotic 
rate of the Lyapunov growth since it describes decay of the quasiparticles. I further approximate Eq.\,\eqref{eq:BS second to last} 
by imposing the on-shell condition, 
\begin{equation}\label{eq:on-shell}
    G^R(k,k_0)(G^R(k,k_0-\omega))^*\approx i\pi\frac{\delta(k_0-\xi_k)}{\omega+i\Im\Sigma_k},
\end{equation} 
which is an analog of the quasiparticle approximation in the derivation of the Boltzmann equation. 
Substituting Eq.\,\eqref{eq:on-shell} into Eq.\,\eqref{eq:BS second to last}, one arrives at the BS equation,
\begin{equation}\label{eq:BS equation final}
    -i\omega f(k,\omega)\approx-\Im\Sigma_kf(k,\omega)+\sum_{n\in\mb{Z}}\sum_{q\in\mr{BZ}}K_{-n}(k,q)f(q,\omega+2nE),
\end{equation}
where $f(k,\omega)=f(k,\xi_k,\omega)$ and the kernel $K$ is defined as $K_{-n}(k,q)=R_n(k,\xi_k,q,\xi_q)/2$.

To solve this equation, I define the ``vector''
\begin{equation}
    f_n(k,\omega)=f(k,\omega+2nE),
\end{equation}
and with this notation Eq.\,\eqref{eq:BS equation final} is written as the Schr\"odinger equation of periodically driven systems,
\begin{equation}\label{eq:Floquet OTOC}
    -i(\omega+2nE)f_n(k,\omega)\approx-\Im\Sigma_kf_n(k,\omega)+\sum_{m\in\mb{Z}}\sum_{q\in\mr{BZ}}K_{-m}(k,q)f_{n+m}(q,\omega).
\end{equation}
I regard $-\Im\Sigma_k$ and $K_{-m}(k,q)$ as a Fourier series of the periodic (matrix) function $\hat{K}(t)$,
\begin{equation}
    \mel*{k}{\hat{K}(t)}{q}\equiv-\delta_{k,q}\Im\Sigma_k+K(k,q,t)=\sum_{n\in\mb{Z}}e^{-i2nEt}\left(-\Im\Sigma_k\delta_{k,q}\delta_{n,0}+K_n(k,q)\right).
\end{equation}
Eq.\,\eqref{eq:Floquet OTOC} is then equivalent to the Schr\"odinger equation $\dv{t}\ket*{f(t)}=\hat{K}(t)\ket*{f(t)}$. The Schr\"odinger equation is formally 
solved by exponentiating the kernel $\hat{K}(t)$, which is expanded as the Magnus series,
\begin{equation}
    \ket*{f(t)}=\mc{T}\exp\left[\int_0^td\tau\hat{K}(\tau)\right]\ket*{f},\,\,\mc{T}\exp\left[\int_0^td\tau\hat{K}(\tau)\right]=\exp\left[\hat{\Omega}_K(t)\right],
\end{equation}
where $\mc{T}$ is the time-ordered product and 
\begin{equation}
    \begin{split}
        \hat{\Omega}_K(t)&=\sum_{n=1}^\infty\hat{\Omega}_n(t)\\
        \hat{\Omega}_1(t)&=\int_0^td\tau\hat{K}(\tau)\\
        \hat{\Omega}_2(t)&=\int_0^td\tau_1\int_0^{\tau_1}d\tau_2[\hat{K}(\tau_1),\hat{K}(\tau_2)]\\
        &\vdots
    \end{split}
\end{equation}
$\hat{\Omega}_1$ is already sufficient for the Lyapunov exponent to be correctly calculated up to the order of $1/N$, as each matrix element of $\hat{K}(t)$ is of order $1/N$. 
If one sets $t=n\pi/E$, $\hat{\Omega}_1$ corresponds to $K_0$. Therefore, Eq.\,\eqref{eq:BS easy main text} is justified.


\section{Derivation of the OTOC of another unstable periodic orbit}
\label{app:OTOC another orbit}
\renewcommand{\theequation}{D\arabic{equation}}
\setcounter{equation}{0}
In this Appendix I aim to derive the OTOC of the periodic orbit $\ket*{\beta}$ [Eq.\,\eqref{eq:another orbit}], 
following closely the derivation of the OTOCs in the previous cases.

The unstable periodic orbit Eq.\,\eqref{eq:another orbit} corresponds to the following 
solution of the equation of motion of the Keldysh action [Eq.\,\eqref{eq:eom Keldysh}],
\begin{equation}
    a_{k\mu\gamma}(t)=0,\,\,b_{k\mu\gamma}(t)=\delta_{k,0}\beta_\mu e^{-iEt}.
\end{equation} 

Following the idea of the sudden quench in Appendix\,\ref{app:OTOC QS}, the interacting term 
$L_{\mr{int}}=\sum_{\gamma=\pm}\gamma L_{\mr{int},\gamma}$ is written as,
\begin{equation}
    \begin{split}
        L_{\mr{int},\gamma}&=-\frac{J\theta(t)}{N^2}\sum_{i=1}^L\sum_{\mu\nu\rho=1}^N
        \left(\beta^*_\mu e^{iEt}+b_{i\mu\gamma}\right)\left(\beta_\mu e^{-iEt}+b_{i\mu\gamma}\right)\left(a_{i\nu\gamma}+a_{i\nu\gamma}^*\right)^2.
    \end{split}
\end{equation}
Assuming $\beta^2\coloneqq N^{-1}\sum_\mu|\beta_\mu|^2$ is sufficiently large, the dominant term in $L_{\mr{int}}$ is 
\begin{equation}\begin{split}
    L_{\mr{int}}&\approx-8J\theta(t)\beta^2\cos^2(Et)\sum_{i=1}^L\sum_{\mu=1}^N\left(a_{i\mu c}+a^*_{i\mu c}\right)\left(a_{i\mu q}+a^*_{i\mu q}\right)\\
    &\equiv-2\mc{J}(t)\sum_{i=1}^L\sum_{\mu=1}^N\left(a_{i\mu c}+a^*_{i\mu c}\right)\left(a_{i\mu q}+a^*_{i\mu q}\right).
\end{split}
\end{equation} 
Note that unlike the initial scar state in Appendix.\,\ref{app:Lyapunov QMBS}, the dominant term is quadratic, which implies that $L_{\mr{int}}$ affects the 
OTOC even at the leading order. Indeed, the quadratic part of the full Lagrangian $L_{\mr{quad}}$ is
{
\begin{equation}
    \begin{split}
        L_{\mr{quad}}&=\frac{1}{2}\sum_{k\in\mr{BZ}}\sum_{\mu=1}^N\begin{pmatrix}
            a_{k\mu c}^*&a_{-k\mu c}
        \end{pmatrix}\begin{pmatrix}
            i\partial_t-i0-\xi_k-\mc{J}(t)&\mc{J}(t)\\
            -\mc{J}(t)&-i\partial_t+i0-\xi_{-k}-\mc{J}(t)
        \end{pmatrix}\begin{pmatrix}
            a_{k\mu q}\\ a^*_{-k\mu q}
        \end{pmatrix}\\
        &+\frac{1}{2}\sum_{k\in\mr{BZ}}\sum_{\mu=1}^N\begin{pmatrix}
            a_{k\mu q}^*&a_{-k\mu q}
        \end{pmatrix}\begin{pmatrix}
            i\partial_t+i0-\xi_k-\mc{J}(t)&\mc{J}(t)\\
            -\mc{J}(t)&-i\partial_t-i0-\xi_{-k}-\mc{J}(t)
        \end{pmatrix}\begin{pmatrix}
            a_{k\mu c}\\ a^*_{-k\mu c}
        \end{pmatrix}\\
        &+2i0\sum_{k\in\mr{BZ}}a^*_{k\mu q}a_{k\mu q}.
    \end{split}
\end{equation}}

Following Appendix.\ref{app:OTOC QS}, I define the Green's function as,
\begin{equation}
    \begin{split}
        i\mc{G}^R(k;t,t')&=\expval{\begin{pmatrix}a_{k\mu c}(t)
            \\ a^*_{-k\mu c}(t)\end{pmatrix}\begin{pmatrix}
            a_{k\mu q}(t')& a^*_{-k\mu q}(t')
            \end{pmatrix}}\\
        i\mc{G}^A(k;t,t')&=\expval{\begin{pmatrix}a_{k\mu q}(t)
            \\ a^*_{-k\mu q}(t)\end{pmatrix}\begin{pmatrix}
            a_{k\mu c}(t')& a^*_{-k\mu c}(t')
            \end{pmatrix}}\\
        i\mc{G}^K(k;t,t')&=\expval{\begin{pmatrix}a_{k\mu c}(t)
                \\ a^*_{-k\mu c}(t)\end{pmatrix}\begin{pmatrix}
                a_{k\mu c}(t')& a^*_{-k\mu c}(t')
                \end{pmatrix}}.
    \end{split}
\end{equation}  
The retarded Green's function defined in the main text (Eq.\,\eqref{eq:another orbit GR}) is 
the $(1,1)$-component of $\mc{G}^R$. 

At the leading order, $\mc{G}^R$ is easily obtained, 
\begin{equation}
    \mc{G}^R(k;t,t')=-i\theta(t)\sum_\alpha\psi_{k\alpha}(t)\psi_{k\alpha}^\dag(t'),\,\,t,t'>0,
\end{equation} 
where $\psi_{k\alpha}$ is the Floquet normal mode satisfying
\begin{equation}
    \left[\sigma^3\left(i\dv{t}+i0^+\right)-\begin{pmatrix}
    \xi_k+\mc{J}(t)&\mc{J}(t)\\\mc{J}(t)&\xi_{-k}+\mc{J}(t)
    \end{pmatrix}\right]\psi_{k\alpha}(t)=0.
\end{equation}
Again, the Floquet normal mode $\psi_{k\alpha}$ satisfies the relation
\begin{equation}
    \psi_{k\alpha}(t)=e^{\lambda_{k\alpha}t}u_{k\alpha}(t),\,\,u_{k\alpha}(t+\pi/E)=u_{k\alpha}(t).
\end{equation}
Therefore, the exponential factor in the normal mode corresponds to the Lyapunov exponent and 
the OTOC at the leading order is expressed as,
\begin{equation}
    C_\beta(t)=\frac{\theta(t)}{NL}\sum_{k\in\mr{BZ}}\left|\sum_\alpha e^{\lambda_{k\alpha}t}\left[u_{k\alpha}(t)u^\dag_{k\alpha}(0)\right]_{11}\right|^2.
\end{equation}  

\section{Derivation of the OTOC with the scar-breaking perturbation}
\label{app:scar-breaking}
In this section I will derive the OTOC of the periodic orbit when the scar-breaking term $V$ is 
added. To this end, I define the following Green's function,
\begin{equation}
    \bm{\mc{G}}(k;t,t')=\begin{pmatrix}
        \mc{G}^K(k;t,t')&\mc{G}^R(k;t,t')\\ \mc{G}^A(k;t,t')&0
    \end{pmatrix},
\end{equation}
where each component is defined as,
\begin{equation}
    \begin{split}
        i\mc{G}^R(k;t,t')&=\expval{
            \begin{pmatrix}
                b_{k\mu c}(t)\\ b^*_{-k\mu c}(t)
            \end{pmatrix}
            \begin{pmatrix}
                b^*_{k\mu q}(t')& b_{-k\mu q}(t')
            \end{pmatrix}
        }\\
        i\mc{G}^A(k;t,t')&=\expval{
            \begin{pmatrix}
                b_{k\mu q}(t)\\ b^*_{-k\mu q}(t)
            \end{pmatrix}
            \begin{pmatrix}
                b^*_{k\mu c}(t')& b_{-k\mu c}^*(t')
            \end{pmatrix}
        }\\
        i\mc{G}^K(k;t,t')&=\expval{
            \begin{pmatrix}
                b_{k\mu c}(t)\\ b^*_{-k\mu c}(t) 
            \end{pmatrix}
            \begin{pmatrix}
                b^*_{k\mu c}(t')& b_{-k\mu c}(t')
            \end{pmatrix}
        }.
    \end{split}
\end{equation}
In the following I denote $\tau^a$ and $\sigma^a$ as the Pauli matrices acting on the Keldysh space and the 
particle-hole space, respectively. 

Note that the quadratic part of the Lagrangian is modified by the perturbation,  
\begin{equation}\begin{split}
    L_{\mr{quad}}&=\sum_{k\in\mr{BZ}}\sum_{\mu=1}^N\begin{pmatrix}
        a^*_{k\mu c}&a_{k\mu q}
    \end{pmatrix}
    \begin{pmatrix}
        0&i\partial_t-i0-\xi_k\\ 
        i\partial_t+i0-\xi_k&2i0
    \end{pmatrix}
    \begin{pmatrix}
        a_{k\mu c}\\a_{k\mu q}
    \end{pmatrix}\\
    &+\frac{1}{2}\sum_{k\in\mr{BZ}}\sum_{\mu=1}^N\begin{pmatrix}
        b_{k\mu c}^*&b_{-k\mu c}
    \end{pmatrix}
    \begin{pmatrix}
        i\partial_t-i0-\xi_k&\theta(t)\epsilon\\
        \theta(t)\epsilon&-i\partial_t+i0-\xi_k
    \end{pmatrix}
    \begin{pmatrix}
        b_{k\mu q}\\ b^*_{-k\mu q}
    \end{pmatrix}\\
    &+\frac{1}{2}\sum_{k\in\mr{BZ}}\sum_{\mu=1}^N\begin{pmatrix}
        b_{k\mu q}^*&b_{-k\mu q}
    \end{pmatrix}
    \begin{pmatrix}
        i\partial_t+i0-\xi_k&\theta(t)\epsilon\\
        \theta(t)\epsilon&-i\partial_t-i0-\xi_k
    \end{pmatrix}
    \begin{pmatrix}
        b_{k\mu q}\\ b_{-k\mu q}^*
    \end{pmatrix}\\
    &+2i0\sum_{k\in\mr{BZ}}\sum_{\mu=1}^Nb_{k\mu q}^*b_{k\mu q}.
\end{split}
\end{equation}
As one can see the discussion below, $L_{\mr{quad}}$ might already indicate the chaos of the periodic orbit 
when the perturbation is strong enough, while for the weak perturbation one has to check the higher orders to 
detect the chaotic dynamics of the OTOC. 
\subsection{The case $|\epsilon|>E$}\label{appsub:e>E}
When the perturbation satisfies the condition $|\epsilon|>E$, the OTOC exhibits the exponential growth just as QS (Appendix.\,\ref{app:OTOC QS}) and the unstable periodic orbit 
(Appendix.\,\ref{app:OTOC another orbit}). 

For $t,t'>0$, the retarded Green's function shows different behaviors depending on 
the sign of $\xi_k^2-\epsilon^2$. If $\xi_k^2>\epsilon^2$, it is  
\begin{equation}\begin{split}\label{eq:GR pert}
    i\bm{\mc{G}}^R(k;t,t')&=\theta(t-t')\frac{e^{-i\sqrt{\xi_k^2-\epsilon^2}t}}{2\sqrt{\xi_k^2-\epsilon^2}}
    \begin{pmatrix}
        \sqrt{\xi_k^2-\epsilon^2}+\xi_k&\epsilon\\
        \epsilon&-\sqrt{\xi_k^2-\epsilon^2}+\xi_k
    \end{pmatrix}\\
    &-\theta(t-t')\frac{e^{i\sqrt{\xi_k^2-\epsilon^2}t}}{2\sqrt{\xi_k^2-\epsilon^2}}
    \begin{pmatrix}
        -\sqrt{\xi_k^2-\epsilon^2}+\xi_k&\epsilon\\
        \epsilon&\sqrt{\xi_k^2-\epsilon^2}+\xi_k
    \end{pmatrix}.
\end{split}
\end{equation}
If $\xi_k^2<\epsilon^2$, it shows the exponential growth,
\begin{equation}
    \begin{split}
        i\bm{\mc{G}}^R(k;t,t')&=\frac{e^{\sqrt{\epsilon^2-\xi_k^2}t}}{2i\sqrt{\epsilon^2-\xi_k^2}}
    \begin{pmatrix}
        i\sqrt{\epsilon^2-\xi_k^2}+\xi_k&\epsilon\\
        \epsilon&-i\sqrt{\epsilon^2-\xi_k^2}+\xi_k
    \end{pmatrix}\\
    &-\frac{e^{-\sqrt{\epsilon^2-\xi_k^2}t}}{2i\sqrt{\epsilon^2-\xi_k^2}}
    \begin{pmatrix}
        -i\sqrt{\epsilon^2-\xi_k^2}+\xi_k&\epsilon\\
        \epsilon&i\sqrt{\epsilon^2-\xi_k^2}+\xi_k
    \end{pmatrix}.
    \end{split}
\end{equation}
Therefore, to the leading order the OTOC shows the exponential growth. 
\subsection{The case $|\epsilon|<E$}\label{appsec:e<E}
Eq.\,\eqref{eq:GR pert} indicates that the quadratic term of the Lagrangian is not enough to obtain the chaos when the perturbation 
satisfies $|\epsilon|<E$, which requires that one should treat the interaction. For the interacting term $L_{\mr{int}}$ [Eq.\,\eqref{eq:L_int}], 
only the first-order correction (Fig.\,\ref{fig:1st order}) survives in the large-$N$ limit,
\begin{equation}\begin{split}\label{eq:1st order perturbed scar}
    \bm{\mc{G}}(k;t,t')&=\bm{\mc{G}}_0(k;t,t')+\int d\tau\Sigma(\tau)\bm{\mc{G}}_0(k;t,\tau)\tau^1\bm{\mc{G}}(k;\tau,t')\\
        &+\int d\tau\Sigma'(\tau)\bm{\mc{G}}_0(k;t,\tau)\tau^1\otimes\sigma^1\bm{\mc{G}}(k;\tau,t'),
\end{split}
\end{equation} 
where the self-energies $\Sigma$ and $\Sigma'$ are defined as 
\begin{equation}
    \begin{split}
        \Sigma(\tau)&=\frac{8J\alpha^2}{L}\sum_{q\in\mr{BZ}}\sinh(2\theta_q)(2\sinh(2\theta_q)+\cosh(2\theta_q))
        \cos^2(E\tau)\sin^2(\sqrt{\xi_q^2-\epsilon^2}\tau)\\
        \Sigma'(\tau)&=\frac{8J\alpha^2}{L}\sum_{q\in\mr{BZ}}\sinh^2(2\theta_q)\cos^2(E\tau)\sin^2(\sqrt{\xi_q^2-\epsilon^2}\tau),
    \end{split}
\end{equation} 
with 
\begin{equation}
    \theta_q=\mr{arctanh}\left(\frac{\xi_q-\sqrt{\xi_q^2-\epsilon^2}}{\epsilon}\right).
\end{equation}
Note that in order to find the early time growth of the OTOC, this treatment is enough, while for later time one needs to solve 
the Hartree equation, i.e., the self-energy should be a functional of the full Green's function. 

The Dyson equation [Eq.\,\eqref{eq:1st order perturbed scar}] indicates that the retarded Green's function should satisfy the following Dyson 
equation,
\begin{equation}\label{eq:dyson perturbed scar}
    \mc{G}^R(k;t,0)=\mc{G}^R_0(k;t,0)+\int_0^td\tau\mc{G}^R_0(k;t,\tau)\left(\Sigma(\tau)+\Sigma'(\tau)\sigma^1\right)\mc{G}^R(k;\tau,0).
\end{equation} 
This equation is solved numerically interatively by updating the Green's function by the equation,
\begin{equation}
    \mc{G}^R_{n+1}(k;t,0)=\mc{G}^R_0(k;t,0)+\int_0^td\tau\mc{G}^R_0(k;t,\tau)\left(\Sigma(\tau)+\Sigma'(\tau)\sigma^1\right)\mc{G}^R_n(k;\tau,0),
\end{equation}  
until the algorithm converges. 

\begin{figure}
    \centering
    \includegraphics[width=0.7\linewidth]{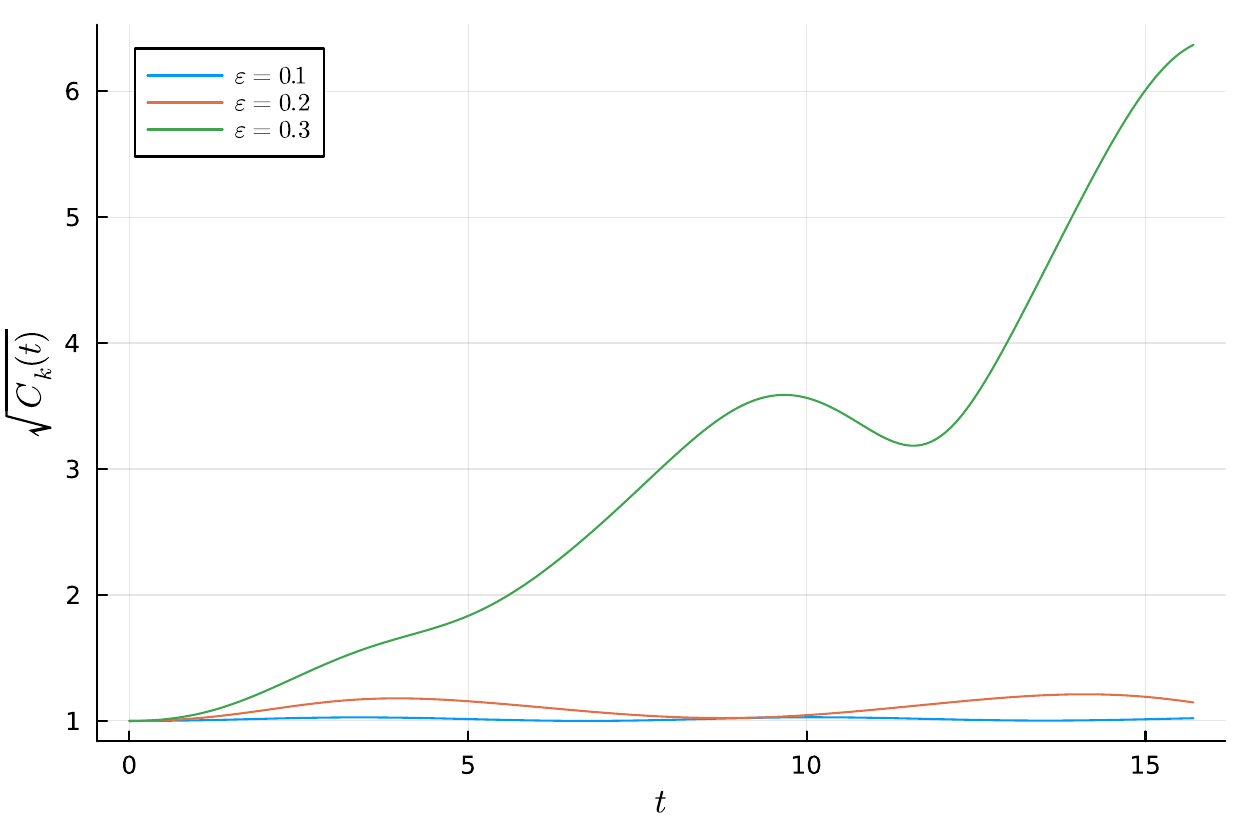}
    \caption{The OTOC $C_k(t)$ at $k=0$ for $\xi_k=-2\cos k+2.5, 8J\alpha^2=3$.}
    \label{fig:otoc_perturb_k0}
\end{figure}

Note that the OTOC is approximately expanded as, 
\begin{equation}
    C_\alpha(t)=\frac{\theta(t)}{N}\sum_{k\in\mr{BZ}}C_k(t).
\end{equation} 
To the leading order $C_k(t)$ is a product of the retarded Green's function. 
I observe numerically that the modes around $k=0$ show the most significant chaotic 
behavior. Furthermore, $C_{k=0}$ in Fig.\,\ref{fig:otoc_perturb_k0} shows that for the small 
$\epsilon$ the growth of the OTOC is extremely slow.

\bibliographystyle{unsrt}
\bibliography{reference}

\end{document}


\title{Supplemental material for PXP paper}
	\author{Keita Arimitsu}
	\affiliation{Department of Physics, ETH Z\"{u}rich, CH-8093 Z\"{u}rich, Switzerland}
	\affiliation{Condensed Matter Theory Group, LSM. NES, Paul Sherrer Institute, Villigen PSI CH-5232, Switzerland}
	\affiliation{Institute of Physics, Ecole Polytechnique F\'ed\'erale de Lausanne (EPFL), CH-1015 Lausanne, Switzerland}
	
\maketitle
\section{Another proof of known exact eigenstates}
There are several exact eigenstates known in PXP model, which was found in Ref.~\cite{Lin2019mps}. They are one zero-energy eigenstate for periodic boundary condition (PBC), and four eigenstates for open boundary condition (OBC). Their wave functions were originally given as matrix product states (MPS), and it was pointed out that a certain basis transformation connect the zero energy eigenstate to the ground state of AKLT model. Later, Ref.~\cite{Shiraishi_2019} gave an elegant proof of the zero energy state, by showing that the basis transformation yields a form $H=\sum_i\mc{P}_ih_i\mc{P}_i+H'$ where $\mc{P}_i$ is a local projection operator. This form is known as Shiraishi-Mori form, which provided a systematic recipe to construct models that violate ETH~\cite{mori_shiraishi}.   

In this section, we provided another proof of these exact eigenstates based on the basis transformation in the main text.
\subsection{Fractionalization}
In order to do so, we introduce a notion of fractionalizing the $N$ spin-$1$ particles into $2N$ spin-$1/2$ particles. We define a map $A_b:\mb{C}^{2\otimes2}\rightarrow\mb{C}^3$ for $\forall b\in\Lambda_{\mr{B}}$:
\begin{equation}
    A_b\coloneqq\ket*{+}_b\bra*{\uparrow\uparrow}_{(b,\mr{L}),(b,\mr{R})}+\ket*{0}_b\frac{1}{\sqrt{2}}\left(\bra*{\uparrow\downarrow}+\bra*{\downarrow\uparrow}\right)_{(b,\mr{L}),(b,\mr{R})}+\ket*{-}_b\bra*{\downarrow\downarrow}_{(b,\mr{L}),(b,\mr{R})},
\end{equation}
where $(b,\mr{L/R})\in\Lambda_{\mr{B}}^{\mr{frac}}\coloneqq\Lambda_{\mr{B}}\times\{\mr{L},\mr{R}\}$ is a lattice site for ``fractional" spin-$1/2$s. Physical meaning of this map is to regard one particle with spin-$1$ as a composite state of two particles with spin-$1/2$, which can be done by discarding the singlet component. We also define a product of such a map as $\msf{A}\coloneqq\prod_{b\in\Lambda_{\mr{B}}}A_b:\mb{C}^{2\otimes2\otimes N}\rightarrow\mb{C}^{3\otimes N}$.  

We can also fractionalize the spin operator with $S=1$ in a sense that for an arbitrary operator $O$ of $S=1$, there exists an operator $O_{\mr{frac}}:\mb{C}^{2\otimes2}\rightarrow\mb{C}^{2\otimes}$ such that $O\msf{A}=\msf{A}O_{\mr{frac}}$. The choice of $O_{\mr{frac}}$ is not unique. For example, both operators $\sqrt{2}\dyad*{\uparrow\uparrow}{\uparrow\downarrow}_{(b,\mr{L}),(b,\mr{R})}$ and $\sqrt{2}\dyad*{\uparrow\uparrow}{\downarrow\uparrow}_{(b,\mr{L}),(b,\mr{R})}$ correspond to $\dyad*{+}{0}$. However, there is a natural choice for $O_{\mr{frac}}$: we first note that for any operator-valued function $F$, it holds that
\begin{equation}\label{eq:operator identity}
    F(\{\bm{S}_b\}_{b\in\Lambda_{\mr{B}}})\msf{A}=\msf{A}F\left(\left\{\bm{S}_{(b,\mr{L})}+\bm{S}_{(b,\mr{R})}\right\}_{b\in\Lambda_{\mr{B}}}\right),
\end{equation}
where $\bm{S}_b$ is a spin operator with $S=1$, i.e., $\bm{S}_b:\mb{C}^3\rightarrow\mb{C}^3$, and $\bm{S}_{(b,\mr{L/R})}$ is a spin operator with $S=1/2$, i.e., $\bm{S}_{(b,\mr{L/R})}:\mb{C}^2\rightarrow\mb{C}^2$. This is physically quite natural, as the spin operator $\bm{S}_b$ is split into a sum of two spin operators $\bm{S}_{(b,\mr{L})}+\bm{S}_{(b,\mr{R})}$.
\subsection{Defining exact eigenstates}
We consider the following states for one zero energy eigenstate $\ket*{\Gamma}$ with PBC and for four eigenstates $\ket*{\Gamma^{\tau\tau'}}$ with OBC:
\begin{equation}\label{eq:|Gamma> in SM}
    \begin{split}
        \ket*{\Gamma}&=\msf{A}\bigotimes_{b\in\Lambda_{\mr{B}}}\frac{1}{\sqrt{2}}\left(\ket*{\uparrow\uparrow}-\ket*{\downarrow\downarrow}\right)_{(b,\mr{R}),(b+1,\mr{L})}\\
        \ket*{\Gamma^{\tau\tau'}}&=\msf{A}\bigotimes_{b\in\Lambda_{\mr{B}}\setminus\{N\}}\frac{1}{\sqrt{2}}\left(\ket*{\uparrow\uparrow}-\ket*{\downarrow\downarrow}\right)_{(b,\mr{R}),(b+1,\mr{L})}\otimes\ket*{\tau}_{(1,\mr{L})}\ket*{\tau'}_{(N,\mr{R})},
    \end{split}
\end{equation}
where $\ket*{\tau}$ and $\ket*{\tau'}$ are either $\ket*{\rightarrow}\coloneqq(\ket*{\uparrow}+\ket*{\downarrow})/\sqrt{2}$ or $\ket*{\leftarrow}\coloneqq(\ket*{\uparrow}-\ket*{\downarrow})/\sqrt{2}$, namely eigenstates of $X$. A pictorial representation of them is described in Fig.~[FIG]. 

We first show that they are elements of $\mc{V}_{\mr{Ryd}}$. To do so, we consider the following state:
\begin{equation}
    \ket*{\gamma^{\sigma\sigma'}_{b,b+1}}\coloneqq \ket{\sigma}_{(b,\mr{L})}\frac{1}{\sqrt{2}}\left(\ket{\uparrow\uparrow}-\ket{\downarrow\downarrow}\right)_{(b,\mr{R}),(b+1,\mr{L})}\ket*{\sigma'}_{b+1,\mr{R}},
\end{equation}
where $\sigma$ and $\sigma'$ are either $\uparrow$ or $\downarrow$, namely eigenstates of  $Z$. Note that $\ket*{\Gamma}$ and $\ket*{\Gamma^{\tau\tau'}}$ are superpositions of $\ket*{\gamma^{\sigma\sigma'}_{b,b+1}}$ in a sense that for $\forall b\in\Lambda_{\mr{B}}$, one can express them as 
\begin{equation}\label{eq:|Gamma> decomposition}
    \begin{split}
         \ket*{\Gamma}&=\msf{A}\sum_{\sigma,\sigma'=\uparrow,\downarrow}c_{\sigma\sigma'}\ket*{\gamma^{\sigma\sigma'}_{b,b+1}}\otimes\ket*{\Xi_{\sigma\sigma'}}\\
         \ket*{\Gamma^{\tau\tau'}}&=\msf{A}\sum_{\sigma,\sigma'=\uparrow,\downarrow}c_{\sigma\sigma'}\ket*{\gamma^{\sigma\sigma'}_{b,b+1}}\otimes\ket*{\Xi^{\tau\tau'}_{\sigma\sigma'}},
    \end{split}
\end{equation}
where $c_{\sigma\sigma'}$ is a coefficient and $\ket*{\Xi_{\sigma\sigma'}}$ and $\ket*{\Xi_{\sigma\sigma'}^{\tau\tau'}}$ are spin-$1/2$ states defined on $(\Lambda_{\mr{B}}\setminus\{b,b+1\})\times\{\mr{L},\mr{R}\}$.
A straightforward calculation yields
\begin{equation}
    \begin{split}
        A_b\otimes A_{b+1}\ket*{\gamma_{b,b+1}^{\uparrow\uparrow}}&=\frac{1}{\sqrt{2}}\ket*{+,+}_{b,b+1}-\frac{1}{2\sqrt{2}}\ket*{0,0}_{b,b+1}\\
        A_b\otimes A_{b+1}\ket*{\gamma_{b,b+1}^{\uparrow\downarrow}}&=\frac{1}{2}\ket*{+,0}_{b,b+1}-\frac{1}{2}\ket*{0,-}_{b,b+1}\\
        A_b\otimes A_{b+1}\ket*{\gamma_{b,b+1}^{\downarrow\uparrow}}&=\frac{1}{2}\ket*{0,+}_{b,b+1}-\frac{1}{2}\ket*{-,0}_{b,b+1}\\
        A_b\otimes A_{b+1}\ket*{\gamma_{b,b+1}^{\downarrow\downarrow}}&=\frac{1}{2\sqrt{2}}\ket*{0,0}_{b,b+1}-\frac{1}{\sqrt{2}}\ket*{-,-}_{b,b+1}.
    \end{split}
\end{equation}
Therefore, combined with Eq.~\eqref{eq:|Gamma> decomposition}, one finds 
\begin{equation}\label{eq:local Pryd}
    \begin{split}
        (1-\dyad{+,-})_{b,b+1}\ket*{\Gamma}&=+\ket*{\Gamma}\\
        (1-\dyad{+,-})_{b,b+1}\ket*{\Gamma^{\tau\tau'}}&=+\ket*{\Gamma^{\tau\tau'}}
    \end{split}
\end{equation}
for $\forall b\in\Lambda_{\mr{B}}$. Since $P_{\mr{Ryd}}$ is a product of $\left(1-\dyad{+,-}\right)_{b,b+1}$ and $\left[(1-\dyad{+,-})_{b,b+1},(1-\dyad{+,-})_{b',b'+1}\right]=0$ for $\forall b,b'\in\Lambda_{\mr{B}}$, namely local constraints commute with each other, we find $P_{\mr{Ryd}}\ket*{\Gamma}=+\ket*{\Gamma}$ and $P_{\mr{Ryd}}\ket*{\Gamma^{\tau\tau'}}=+\ket*{\Gamma^{\tau\tau'}}$. This means $\ket*{\Gamma},\ket*{\Gamma^{\tau\tau'}}\in\mc{V}_{\mr{Ryd}}$.
\subsection{A detailed proof of exact eigenstates}
As $P_{\mr{Ryd}}\ket*{\Gamma}=+\ket*{\Gamma}$, it holds that
\begin{equation}
    H\ket*{\Gamma}=HP_{\mr{Ryd}}\ket*{\Gamma}=P_{\mr{Ryd}}H\ket*{\Gamma}=P_{\mr{Ryd}}\left(H_{\mr{Z}}+H_1\right)\ket*{\Gamma}.
\end{equation}
Here we use $\left[P_{\mr{Ryd}},H\right]=0$ and $P_{\mr{Ryd}}\dyad*{+,-}{\psi}_{b,b+1}=0$ for $\forall\ket{\psi}\in\mb{C}^{3\otimes2}$. Also, Eq.~\eqref{eq:local Pryd} implies that $\dyad*{\psi}{+,-}_{b,b+1}\ket*{\Gamma}=0$ for $\forall b\in\Lambda_{\mr{B}}$. Thus, we obtain $H\ket*{\Gamma}=P_{\mr{Ryd}}H_{\mr{Z}}\ket*{\Gamma}$.
The same relation holds for $\ket*{\Gamma^{\tau\tau'}}$ as well. With the operator identity in Eq.~\eqref{eq:operator identity}, we find
\begin{equation}\begin{split}
    H_{\mr{Z}}\ket{\Gamma}&=\msf{A}\left(\sqrt{2}\sum_{b\in\Lambda_{\mr{B}}}\left(\frac{1}{2}X_{(b,\mr{L})}+\frac{1}{2}X_{(b,\mr{R})}\right)\right)\bigotimes_{b'\in\Lambda_{\mr{B}}}\frac{1}{\sqrt{2}}\left(\ket{\uparrow\uparrow}-\ket{\downarrow\downarrow}\right)_{(b',\mr{R}),(b'+1,\mr{L})}\\
    &=\msf{A}\left(\sqrt{2}\sum_{b\in\Lambda_{\mr{B}}}\left(\frac{1}{2}X_{(b,\mr{R})}+\frac{1}{2}X_{(b+1,\mr{L})}\right)\right)\bigotimes_{b'\in\Lambda_{\mr{B}}}\frac{1}{\sqrt{2}}\left(\ket{\uparrow\uparrow}-\ket{\downarrow\downarrow}\right)_{(b',\mr{R}),(b'+1,\mr{L})}\\&=0,
    \end{split}
\end{equation}
where PBC is used in the second line. For $\ket*{\Gamma^{\tau\tau'}}$, we find
\begin{equation}
    \begin{split}
        H_{\mr{Z}}\ket*{\Gamma^{\tau\tau'}}&=\msf{A}\left(\sqrt{2}\sum_{b\in\Lambda_{\mr{B}}}\left(\frac{1}{2}X_{(b,\mr{L})}+\frac{1}{2}X_{(b,\mr{R})}\right)\right)\bigotimes_{b'\in\Lambda_{\mr{B}}\setminus\{N\}}\frac{1}{\sqrt{2}}\left(\ket*{\uparrow\uparrow}-\ket*{\downarrow\downarrow}\right)_{(b',\mr{R}),(b'+1,\mr{L})}\otimes\ket*{\tau}_{(1,\mr{L})}\ket*{\tau'}_{(N,\mr{R})}\\
        &=\msf{A}\frac{1}{\sqrt{2}}\left(X_{(1,\mr{L})}+X_{(N,\mr{R})}\right)\bigotimes_{b'\in\Lambda_{\mr{B}}\setminus\{N\}}\frac{1}{\sqrt{2}}\left(\ket*{\uparrow\uparrow}-\ket*{\downarrow\downarrow}\right)_{(b',\mr{R}),(b'+1,\mr{L})}\otimes\ket*{\tau}_{(1,\mr{L})}\ket*{\tau'}_{(N,\mr{R})}.
    \end{split}
\end{equation}
As $\ket{\tau}$ and $\ket*{\tau'}$ are eigenstates of $X$, $\ket*{\Gamma^{\tau\tau'}}$ is also an eigenstate of $H_{\mr{Z}}$.
\subsection{Matrix product state representation}
In Ref.~\cite{Lin2019mps}, the zero energy eigenstate for PBC and four eigenstates for OBC are given as matrix product states (MPS). We re-derive MPS representation of $\ket*{\Gamma}$ and $\ket*{\Gamma^{\tau\tau'}}$ and show an equivalence between the states considered in Ref.~\cite{Lin2019mps} and the states in Eq.~\eqref{eq:|Gamma> in SM}. To do so, we define a bond variable $\rket{\alpha}_b$ 
for $b\in\Lambda_{\mr{B}}$ such that
\begin{equation}
    \rket{1}_b\coloneqq\ket*{\uparrow\uparrow}_{(b-1,\mr{R}),(b,\mr{L})},\,\,
    \rket{2}_b\coloneqq\ket*{\downarrow\downarrow}_{(b-1,\mr{R}),(b,\mr{L})}.
\end{equation}
$\ket*{\Gamma}$ is re-written as
\begin{equation}
    \begin{split}
        \ket*{\Gamma}&=\msf{A}\bigotimes_{b\in\Lambda_{\mr{B}}}\frac{1}{\sqrt{2}}\left(\rket{1}-\rket{2}\right)_b.
    \end{split}
\end{equation}
From this expression, we can find MPS representation for $\ket*{\Gamma}$: a straightforward calculation yields
\begin{equation}\begin{split}
    A_b\frac{1}{\sqrt{2}}\left(\rket{1}-\rket{2}\right)_b\otimes\frac{1}{\sqrt{2}}\left(\rket{1}-\rket{2}\right)_{b+1}&=\msf{A}_{1,1}^+\ket*{\uparrow}_{(b-1,\mr{R})}\ket*{+}_b\ket*{\uparrow}_{(b+1,\mr{L})}+\msf{A}_{1,2}^0\ket*{\uparrow}_{(b-1,\mr{R})}\ket*{0}_b\ket*{\downarrow}_{(b+1,\mr{L})}\\
    &-\msf{A}_{2,1}^0\ket*{\downarrow}_{(b-1,\mr{R})}\ket*{0}^b\ket*{\uparrow}_{(b+1,\mr{L})}-\msf{A}_{2,2}^-\ket*{\downarrow}_{(b-1,\mr{R})}\ket*{-}_b\ket*{\downarrow}_{(b+1,\mr{L})},
    \end{split}
\end{equation}
where 
\begin{equation}
\msf{A}_{1,1}^+\coloneqq\frac{1}{2},\,\,\msf{A}^0_{1,2}\coloneqq-\frac{1}{2\sqrt{2}},\,\,\msf{A}^0_{2,1}\coloneqq\frac{1}{2\sqrt{2}},\,\,\msf{A}^-_{2,2}\coloneqq-\frac{1}{2}.
\end{equation}
Therefore, $\ket*{\Gamma}$ is written as
\begin{equation}
    \ket*{\Gamma}=\sum_{\{\sigma_i\}_{i=1}^N\in\{\pm,0\}^{\otimes N}}\mr{Tr}\left[\msf{A}^{\sigma_1}\cdots\msf{A}^{\sigma_{N}}\right]\ket*{\sigma_1\cdots\sigma_N},
\end{equation}
where 
\begin{equation}
    \msf{A}^+\coloneqq\frac{1}{2\sqrt{2}}\begin{pmatrix}\sqrt{2} & 0\\0&0\end{pmatrix},\,\,\msf{A}^0\coloneqq\frac{1}{2\sqrt{2}}\begin{pmatrix}0&-1\\1&0\end{pmatrix},\,\,\msf{A}^-\coloneqq\frac{1}{2\sqrt{2}}\begin{pmatrix}0&0\\0&-\sqrt{2}\end{pmatrix}.
\end{equation}
Up to an irrelevant normalization factor, this is the same as the zero energy state in Ref.~\cite{Lin2019mps}.

A similar consideration leads to MPS representation of $\ket*{\Gamma^{\tau\tau'}}$:
\begin{equation}
    \begin{split}
        \ket*{\Gamma^{\tau\tau}}=\sum_{\{\sigma_i\}_{i=1}^N\in\{\pm,0\}^{\otimes N}}\left(\msf{v}_\tau^{\sigma_1}\right)^{\mr{T}}\msf{A}^{\sigma_2}\cdots\msf{A}^{\sigma_{N-1}}\msf{v}_{\tau'}^{\sigma_N}\ket*{\sigma_1\cdots\sigma_N},
    \end{split}
\end{equation}
where 
\begin{equation}
    \begin{split}
        \msf{v}_\rightarrow^
    \end{split}
\end{equation}
\section{Another set of trial wave functions}
In the main text, we consider the trial wave function $\ket*{S_n}$ that is viewed as the eigenstates of $H_{\mr{Z}}$ with the maximal spin number projected onto the constraint subspace. However, one can also construct another set of trial wave functions based on $\ket*{\Gamma}$. We consider
\begin{equation}\label{eq:another Sn}
    \ket*{S'_n}\coloneqq P_{\mr{Ryd}}\left(J^+\right)^n\ket*{\Gamma},
\end{equation}
where $J^\pm$ is defined in the main text. We numerically observe that squared overlap $|\innerproduct*{S'_n}{\mc{S}^{\mr{PXP}}_n}|^2$ is roughly as large as $|\innerproduct*{S_n}{\mc{S}^{\mr{PXP}}_n}|^2$. While these states are not good approximation to the scar states of perturbed PXP model, $\ket*{S'_n}$ provides a good estimate of the energy of $\ket*{S'_1}$. Using MPS representation, $\ket*{S'_1}$ is written as
\begin{equation}
    \ket*{S'_1}=\sum_{b\in\Lambda_{\mr{B}}}
\end{equation}
\section{Estimate of the optimal coefficient of the perturbation}
We have obtained the optimal coefficient of the perturbation $\delta H(\lambda)$ in the main text. Here we derive it in a detail. Our idea is to minimize $\sum_n\norm*{(H_1+\delta H(\lambda))\ket*{\wtil{S}_n}}^2$ with respect to $\lambda$, instead of $\sum_n\norm*{P_{\mr{Ryd}}(H_1+\delta H(\lambda))\ket*{S_n}}^2$, assuming that they are not so different. 
\subsection{The scar states}
Our trial wave function $\ket{S_n}$ for the scar states is given as follows,
\begin{equation}\begin{split}
    \ket{S_n}&=P_{\mr{Ryd}}\ket*{\wtil{S}_n}=P_{\mr{Ryd}}\left(J^-\right)^{N-n}\bigotimes_{b\in\Lambda_{\mr{B}}}\ket*{\what{+}}_b
    \\
    J^\pm&=\sqrt{2}\sum_{b\in\Lambda_{\mr{B}}}\left(\dyad*{\what{\pm}}{\what{0}}+\dyad*{\what{0}}{\what{\mp}}\right)_b.
    \end{split}
\end{equation}
We split the collective spin-raising (lowering) operator as
\begin{equation}\begin{split}
    J^\pm&=J_{b,b+1}^\pm+J_{\Lambda_{\mr{B}}\setminus\{b,b+1\}}^\pm\\
    J_{b,b+1}^\pm&\coloneqq\sqrt{2}\left(\dyad*{\what{\pm}}{\what{0}}+\dyad*{\what{0}}{\what{\mp}}\right)_b+\sqrt{2}\left(\dyad*{\what{\mp}}{\what{0}}+\dyad*{\what{0}}{\what{\mp}}\right)_{b+1}\\
    J^\pm_{\Lambda_{\mr{B}}\setminus\{b,b+1\}}&\coloneqq\sqrt{2}\sum_{b'\in\Lambda_{\mr{B}}\setminus\{b,b+1\}}\left(\dyad*{\what{\pm}}{\what{0}}+\dyad*{\what{0}}{\what{\mp}}\right)_{b'}.
    \end{split}
\end{equation}
Using these operators, we can write $\ket*{\wtil{S}_{N-n}}$ with $n\geq3$ as
\begin{equation}\begin{split}
    \ket*{\wtil{S}_{N-n}}&=\sum_{k=0}^4\binom{n}{k}\left(J^-_{b,b+1}\right)^k\ket*{\what{T}_{2,2}}_{b,b+1}\otimes\left(J^-_{\Lambda_{\mr{B}}\setminus\{b,b+1\}}\right)^{n-k}\ket*{\what{T}_{N-2,N-2}}_{\Lambda_{\mr{B}}\setminus\{b,b+1\}}\\
    &=c\ket*{\what{T}_{2,2}}_{b,b+1}\otimes\ket*{\what{T}_{N-2,N-n-2}}_{\Lambda_{\mr{B}}\setminus\{b,b+1\}}+2c\sqrt{\frac{n}{-n+2N-3}}\ket*{\what{T}_{2,1}}_{b,b+1}\otimes\ket*{\what{T}_{N-2,N-n-1}}_{\Lambda_{\mr{B}}\setminus\{b,b+1\}}\\
    &+c\sqrt{\frac{3n!(-n+2N-4)!}{2(n-2)!(-n+2N-2)!}}\ket*{\what{T}_{2,0}}_{b,b+1}\otimes\ket*{\what{T}_{N-2,N-n}}_{\Lambda_{\mr{B}}\setminus\{b,b+1\}}\\
    &+c\sqrt{\frac{n!(-n+2N-4)!}{6(n-3)!(-n+2N-1)!}}\ket*{\what{T}_{2,-1}}_{b,b+1}\otimes\ket*{\what{T}_{N-2,N-n+1}}_{\Lambda_{\mr{B}}\setminus\{b,b+1\}}\\
    &+c\sqrt{\frac{n!(-n+2N-4)!}{(n-4)!(-n+2N)!}}\ket*{\what{T}_{2,-2}}_{b,b+1}\otimes\ket*{\what{T}_{N-2,N-n+2}}_{\Lambda_{\mr{B}}\setminus\{b,b+1\}},
    \end{split}
\end{equation}
where $c\coloneqq\prod_{M=N-n-1}^{N-2}\sqrt{(N-2)(N-1)-M(M-1)}$. Here, $\ket*{\what{T}_{S,M}}$ is a composite state with total spin $S$ and $S^x=M$. When $N$ is sufficiently large, one can approximate $\ket*{\wtil{S}_n}$ as
\begin{equation}\label{eq:|Sn> approximate}
    \begin{split}
        \frac{1}{c}\ket*{\wtil{S}_{N-n}}&\cong \ket*{\what{T}_{2,2}}_{b,b+1}\otimes\ket*{\what{T}_{N-2,N-n-2}}_{\Lambda_{\mr{B}}\setminus\{b,b+1\}}+2\sqrt{\frac{s}{2-s}}\ket*{\what{T}_{2,1}}_{b,b+1}\otimes\ket*{\what{T}_{N-2,N-n-1}}_{\Lambda_{\mr{B}}\setminus\{b,b+1\}}\\
        &+\sqrt{\frac{3}{2}}\frac{s}{2-s}\ket*{\what{T}_{2,0}}_{b,b+1}\otimes\ket*{\what{T}_{N-2,N-n}}_{\Lambda_{\mr{B}}\setminus\{b,b+1\}}+\sqrt{\frac{s^3}{6(2-s)^3}}\ket*{\what{T}_{2,-1}}_{b,b+1}\otimes\ket*{\what{T}_{N-2,N-n+1}}_{\Lambda_{\mr{B}}\setminus\{b,b+1\}}\\
        &+\frac{1}{12}\frac{s^2}{(2-s)^2}\ket*{\what{T}_{2,-2}}_{b,b+1}\otimes\ket*{\what{T}_{N-2,N-n+2}}_{\Lambda_{\mr{B}}\setminus\{b,b+1\}},
    \end{split}
\end{equation}
where $s\coloneqq n/N$. Each coefficient is plotted in Fig.~[FIG]. When $n$ is small, the dominant contribution in $\ket*{\wtil{S}_{N-n}}$ is $\ket*{\what{T}_{2,2}}_{b,b+1}$, but as $n$ increases $\ket*{\what{T}_{2,1}}_{b,b+1}$ and $\ket*{\what{T}_{2,0}}_{b,b+1}$ become dominant. For later argument, we define the reduced density matrix $\rho_S$ for the states $\ket*{\wtil{S}_n}$ as 
\begin{equation}
    \rho_S\coloneqq\frac{1}{N}\sum_{n=1}^N\mr{Tr}_{\Lambda_{\mr{B}}\setminus\{b,b+1\}}\frac{1}{\norm*{\ket*{\wtil{S}_n}}^2}\dyad*{\wtil{S}_n},
\end{equation}
where $\mr{Tr}_{\Lambda_{\mr{B}}\setminus\{b,b+1\}}$ is partial trace on $\Lambda_{\mr{B}}\setminus\{b,b+1\}$. We can approximately obtain $\rho_S$ using Eq.~\eqref{eq:|Sn> approximate} as
\begin{equation}
    \begin{split}
        \rho_S&\cong\frac{1}{N}\sum_{s=1/N}^{1}\frac{1}{Z(s)}\left(P^{(2,2)}_{b,b+1}+\frac{4s}{2-s}P^{(2,1)}_{b,b+1}+\frac{3s^2}{2(2-s)^2}P^{(2,0)}_{b,b+1}+\frac{s^3}{6(2-s)^3}P^{(2,-1)}_{b,b+1}+\frac{s^4}{144(2-s)^4}P^{(2,-2)}_{b,b+1}\right)\\
        Z(s)&\coloneqq1+\frac{4s}{2-s}+\frac{3s^2}{2(2-s)^2}+\frac{s^3}{6(2-s)^3}+\frac{s^4}{144(2-s)^4},
    \end{split}
\end{equation}
where $P_{b,b+1}^{(S,M)}\coloneqq\dyad*{\what{T}_{S,M}}_{b,b+1}$. Note that we can approximate $\rho_S$ further by replacing $N^{-1}\sum_{s=1/N}^1\rightarrow\int_0^1ds$.
\begin{figure}
    \centering
    \includegraphics[width=.5\textwidth]{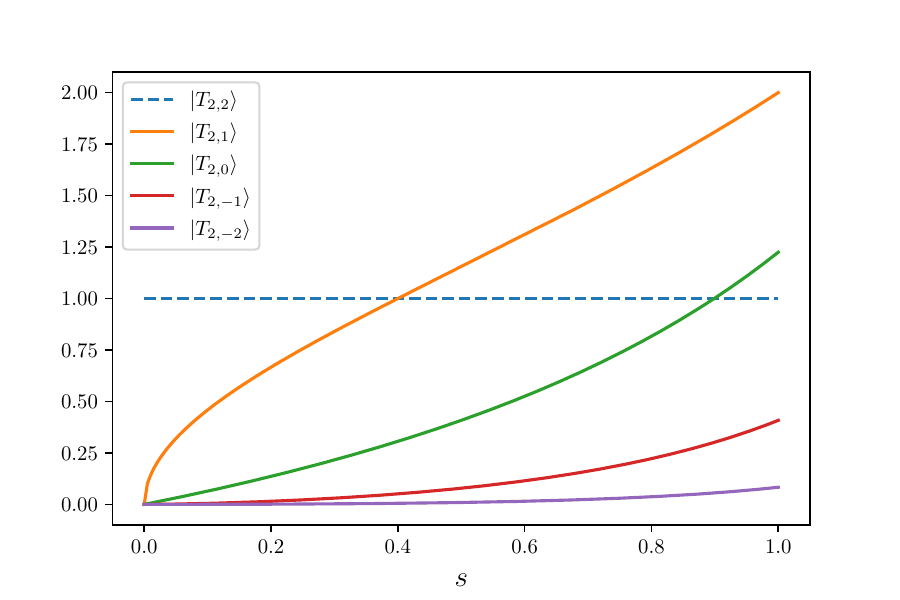}
    \caption{Coefficients of $\ket*{\what{T}_{2,M}} (-2\leq M\leq 2)$ in Eq.~\eqref{eq:|Sn> approximate} as functions of $s$.}
    \label{fig:coeffs}
\end{figure}
\subsection{The estimate of the optimal coefficient}
As shown in the main text, we have found 
\begin{equation}
    \begin{split}
        (H+\delta H(\lambda))\ket*{S_n}&=P_{\mr{Ryd}}\left(H_{\mr{Z}}+H_{\mr{rem}}(\lambda)\right)\ket*{\wtil{S}_n}\\
        H_{\mr{rem}}(\lambda)&=\sum_{b\in\Lambda_{\mr{B}}}h_{b,b+1}(\lambda)\\
        h_{b,b+1}(\lambda)&=\frac{-1+2\lambda}{\sqrt{6}}\left(\ket*{+,0}+\ket*{0,-}\right)\bra*{T_{2,0}}_{b,b+1}+\frac{\lambda}{\sqrt{2}}\ket*{0,0}\left(\bra*{T_{2,1}}+\bra*{T_{2,-1}}\right)_{b,b+1}.
    \end{split}
\end{equation}
We find
\begin{equation}
    h_{b,b+1}^\dag(\lambda)h_{b,b+1}(\lambda)=\frac{(-1+2\lambda)^2}{3}\dyad*{T_{2,0}}_{b,b+1}+\frac{\lambda^2}{2}\left(\ket*{T_{2,1}}+\ket*{T_{2,-1}}\right)\left(\bra*{T_{2,1}}+\bra*{T_{2,-1}}\right)_{b,b+1}.
\end{equation}
As $\ket*{T_{2,0}}=\sqrt{3/8}\ket*{\what{T}_{2,2}}-1/2\ket*{\what{T}_{2,0}}+\sqrt{3/8}\ket*{\what{T}_{2,-2}}$ and $\ket*{T_{2,1}}+\ket*{T_{2,-1}}=\ket*{\what{T}_{2,2}}-\ket*{\what{T}_{2,-2}}$, we find
\begin{equation}
    \begin{split}
        \mr{Tr}_{\{b,b+1\}}\rho_Sh^\dag_{b,b+1}(\lambda)h_{b,b+1}(\lambda)&=\frac{(-1+2\lambda)^2}{3}\left(\frac{3}{8}\expval*{\rho_S}{\what{T}_{2,2}}+\frac{1}{4}\expval*{\rho_S}{\what{T}_{2,0}}+\frac{3}{8}\expval*{\rho_S}{\what{T}_{2,-2}}\right)\\
        &+\frac{\lambda^2}{2}\left(\expval*{\rho_S}{\what{T}_{2,2}}+\expval*{\rho_S}{\what{T}_{2,-2}}\right)\\
        &\cong\frac{1}{N}\sum_{s=1/N}^1\left[\frac{(-1+2\lambda)^2}{8Z(s)}\left(1+\frac{s^2}{(2-s)^2}+\frac{s^4}{144(2-s)^4}\right)+\frac{\lambda^2}{2Z(s)}\left(1+\frac{s^4}{144(2-s)^4}\right)\right]\\
        &\cong\int_0^1ds\left[\frac{(-1+2\lambda)^2}{8Z(s)}\left(1+\frac{s^2}{(2-s)^2}+\frac{s^4}{144(2-s)^4}\right)+\frac{\lambda^2}{2Z(s)}\left(1+\frac{s^4}{144(2-s)^4}\right)\right],
    \end{split}
\end{equation}
where $\mr{Tr}_{\{b,b+1\}}$ is partial trace on $\{b,b+1\}$. This function becomes smallest at $\lambda=\alpha/(2(\alpha+\beta))$, where
\begin{equation}
    \begin{split}
        \alpha&\coloneqq\int_0^1ds\frac{1}{Z(s)}\left(1+\frac{s^2}{(2-s)^2}+\frac{s^4}{144(2-s)^4}\right)\cong0.5350\\
        \beta&\coloneqq\int_0^1ds\frac{1}{Z(s)}\left(1+\frac{s^4}{144(2-s)^4}\right)\cong0.4739.
    \end{split}
\end{equation}
Thus, we numerically find the optimal coefficient $\lambda\cong0.2651$.
\section{A possible way to generate a sequence of Hermitian perturbations}

\bibliographystyle{unsrt}
\bibliography{reference}